\documentclass{IEEEtran}
\IEEEoverridecommandlockouts
\usepackage{cite}
\usepackage{amsmath,amssymb,amsfonts}
\usepackage{algorithmic}
\usepackage{graphicx}
\usepackage{textcomp}
\usepackage{xcolor}
\usepackage{setspace}
\def\BibTeX{{\rm B\kern-.05em{\sc i\kern-.025em b}\kern-.08em
		T\kern-.1667em\lower.7ex\hbox{E}\kern-.125emX}}
\usepackage{subfigure}
\newtheorem{my_theorem}{Theorem}

\newtheorem{my_lemma}{Lemma}

\usepackage{enumitem}
\newcommand*{\J}{\jmath}%

\title{Performance of Hybrid THz and Multi-Antenna RF System with Diversity Combining}

\author{Pranay Bhardwaj,~\IEEEmembership{Graduate Student Member,~IEEE} and S.~M.~Zafaruddin,~\IEEEmembership{Senior Member,~IEEE}
	
	\thanks{ A part of this paper containing an approximate analysis on the THz link mixed with an equal gain combining (EGC) RF receiver has been submitted for presentation in the 2022 IEEE 95th Vehicular Technology Conference: VTC2022-Spring to be held in Helsinki, Finland 19-22 June 2022.}
	
	\thanks{Pranay Bhardwaj (p20200026@pilani.bits-pilani.ac.in) and S.~M.~Zafaruddin (syed.zafaruddin@pilani.bits-pilani.ac.in) are with the Department of Electrical and Electronics Engineering, Birla Institute of Technology and Science, Pilani, Pilani-333031, Rajasthan, India.}
	
	\thanks{This work was supported	in part by the Science and Engineering Research Board (SERB), Department of Science and Technology (DST), Government of India, through the Mathematical Research Impact Centric Support (MATRICS) scheme under Grant MTR/2021/000890, and Start-up Research Grant SRG/2019/002345.}	
}

\begin{document}
	\maketitle 	
\begin{abstract}
	Recent studies  investigate single-antenna radio frequency (RF) systems mixed with terahertz (THz) wireless communications. This paper considers a two-tier system  of THz for backhaul and multiple-antenna assisted RF for the access network. We analyze the system performance by employing both selection combining (SC) and maximal ratio combining (MRC) receivers for the RF link integrated with the THz using the fixed-gain  amplify and forward (AF) protocol. We develop the  probability density function (PDF) and cumulative distribution function (CDF) of the end-to-end signal-to-noise (SNR) of the dual-hop system over independent and non-identically distributed (i.ni.d.)  $\alpha$-$\mu$ fading channels with a statistical model for misalignment errors in the THz wireless link.  We use the derived statistical results to develop analytical expressions of the outage probability, average bit error rate (BER), and ergodic capacity for the performance assessment of the considered system. We develop diversity order of the system using  asymptotic analysis in the high SNR region, demonstrating the scaling of system performance with the number of antennas. We use computer simulations to show the effect of system and channel parameters on  the performance of the hybrid THz-RF link with multi-antenna diversity schemes.
\end{abstract}

	
\section{Introduction}
The advent of the terahertz (THz) spectrum for communications has sparked research to realize an all wireless link consisting of access network over radio frequency (RF) and backhaul/fronthaul in the THz band \cite{Koenig_2013_nature, Boulogeorgos_2018_beyond5g, Elayan_2019}. The THz wireless can be a potential alternative for the fiber, especially for the ubiquitous connectivity in difficult terrain. Moreover, THz signals are immune to atmospheric turbulence as observed in free-space optical (FSO) communications.  A hybrid network consisting of THz and RF technologies can be a potential architecture for the next generation wireless systems. However, THz transmission suffers from higher path loss due to the molecular absorption at a higher frequency, signal fading, and random pointing errors caused by misalignment between transmitting and receiving antennas \cite{Kokkoniemi_2018, KOKKONIEMI2020,Priebe_2011, Olutayo_2020}. It is desirable to analyze the performance of THz wireless systems when integrated with the matured RF technology.

Recent works investigate the performance of  relay-assisted THz transmissions \cite{Alexandros2020_outage, ants2021_pranay,  Giorgos2020, Xia_2017, Rong_2017,Abbasi_2017,Mir2020,Liang_2021_THz_AF, huang2021, bhardwaj2021_multihop} and  mixed RF-THz systems \cite{Boulogeorgos_Error,Pranay_2021_TVT,Pranay_2021_Letters}. The outage probability for a decode-and-forward (DF) based dual-hop THz-THz system was analyzed in \cite{Alexandros2020_outage, ants2021_pranay}.  A multi-relay selection strategy to mitigate the antenna misalignment and channel fading was proposed in \cite{Giorgos2020}. The authors in \cite{Xia_2017} suggested a relaying scheme to optimize the network	throughput  for THz-band wireless  communications.  Amplify-and-forward (AF) relaying was studied for In-Vivo nano communication at THz frequencies \cite{Rong_2017,Abbasi_2017}. The authors in \cite{Mir2020} analyzed the performance of a two-way AF relaying for THz massive multi-input multiple-output
systems. The authors in \cite{Liang_2021_THz_AF} analyzed the performance of a dual-hop THz system using fixed-gain AF relaying. The authors in \cite{huang2021} proposed a deep reinforcement learning (DRL) based multi-hop reconfigurable intelligent surface (RIS) assisted THz transmissions to extend the coverage range.  Recently, we presented the performance of a multihop THz communication system under various channel impairments in \cite{bhardwaj2021_multihop}.

It should be mentioned that the mixed system of  FSO interfaced with multi-antenna/multi-user RF system has been extensively investigated \cite{Yang_2015_multiantenna_RF, Malek_2016_multiantenna_RF,Yang2017,Trigui_2019_multiantenna_RF,Cherif_2017_multiantenna_FSO,Zhang_2020_multiantenna_receiver}.  The authors in \cite{Yang_2015_multiantenna_RF} presented the performance of a variable-gain AF based mixed system with transmit diversity for the multi-antenna source on the RF over Rayleigh fading and opportunistic selection on the multi-aperture FSO link. The physical layer security issues were discussed in \cite{Malek_2016_multiantenna_RF} considering RF/FSO as Nakagami-$m$/gamma-gamma distributed using variable gain with opportunistic user scheduling and multi-antenna diversity techniques (SC and MRC) at the relay. A unified performance exploiting multi-user diversity over $\eta$-$\mu$ RF channel integrated with the FSO link using fixed gain relaying was presented in \cite{Yang2017}. The authors in \cite{Cherif_2017_multiantenna_FSO} developed asymptotic analysis for the mixed FSO/RF system with the best user selection from multiple users on RF channels with  Nakagami-$m$ fading in the presence of co-channel signal interference. The authors in \cite{Trigui_2019_multiantenna_RF} studied the performance of a mixed FSO and RF  multi-user network with aperture selection and opportunistic user scheduling over shadowed $\kappa$-$\mu$ RF channels in the presence of Poisson field interference. A multiple relay setup was considered in \cite{Zhang_2020_multiantenna_receiver}, where the best relay was selected to forward the data from a source over Nakagami-$m$ RF channel to a multi-aperture FSO detector.

There is limited research on the RF-THz hybrid network\cite{Boulogeorgos_Error,Pranay_2021_TVT,Pranay_2021_Letters}. The authors in \cite{Boulogeorgos_Error} considered the DF relaying protocol to facilitate communication for a heterogeneous THz-RF system. They analyzed the outage probability and average bit-error-rate of the mixed system by developing probability density function (PDF) and cumulative distribution function (CDF) of the signal-to-noise ratio (SNR) considering independent and identically distributed (i.i.d.) $\alpha$-$\mu$ fading for both the links.  It is well known that  $\alpha$-$\mu$  is a generalized model which includes Weibull, negative exponential, popular Nakagami-$m$, and Rayleigh distribution as a particular case to model short-term fading at RF frequencies \cite{Yacoub_alpha_mu}.  Further, $\alpha$-$\mu$  has been experimentally validated distribution to model the channel fading in the THz band at $152$ \mbox{GHz} \cite{Papasotiriou2021}. In \cite{Pranay_2021_TVT}, we generalized the performance analysis of the THz-RF link considering asymmetrical $\alpha$-$\mu$  fading for both the links with real-valued $\mu$.  Recently, we employed fixed-gain AF relaying for RF-THz transmissions with $\alpha$-$\kappa$-$\mu$ shadowed fading channel for the access link over RF and  $\alpha$-$\mu$ fading with nonzero boresight pointing error for  THz transmissions  \cite{Pranay_2021_Letters}. While scanning the aforementioned and related research for the hybrid-RF-THz network, we found that a single-antenna RF system has been considered. In general, RF  base-station/access points employ multiple antennas to harness spatial diversity to deal with the signal fading and thus enhance the system performance. To the best of the author's knowledge, the performance analysis of a fixed-gain AF relaying for multi-antenna RF  over $\alpha$-$\mu$ channel fading mixed with high frequency technologies (such as FSO and THz) is not available in the literature. It requires novel approaches to derive statistical results of the end-to-end system in a closed-form when the diversity combining technique is applied at the RF receiver and integrated with the THz link over generalized fading channels.

In this paper, we analyze the performance of a two-tier  system consisting of a single-antenna THz link and a multiple antenna receiver system for RF transmission in an uplink wireless network. In the following, we summarize the main contributions of the paper:

\begin{itemize}
	\item  We  develop PDF and CDF of the end-to-end SNR of the hybrid system by employing both selection combining (SC) and maximal ratio combining (MRC) diversity techniques for the multi-antenna equipped RF receiver over independent and non-identically distributed (i.ni.d) $\alpha$-$\mu$ fading channels with a statistical model for the antenna misalignment  in the THz link.
	
	\item The derived statistical results are presented in terms of  Fox's H-function  due to the manifestation of  fixed gain AF relaying to integrate the RF and THz links with i.ni.d  generalized channel characteristics. 
	\item   We analyze the system performance by deriving analytical expressions of the outage probability, average bit error rate (BER), and ergodic capacity. 
	
	\item  We develop diversity order of the system using  asymptotic analysis in the high SNR region, which demonstrates the effect and channel  parameters and scaling of system performance with the number of antennas. The diversity order provides useful  engineering insights on the deployment issues of  integrating multi-antenna RF base station/access point with the THz backhaul.
	
	\item We use computer simulations to show the effect of system  parameters on  the performance of the considered  hybrid THz-RF  with   multi-antenna diversity schemes.
\end{itemize}

The paper is organized as follows: We describe the system model for the dual-hop multi-antenna RF and THz transmission with PDF and CDF of individual links in Section II. We provide statistical results and performance analysis with receiver selection diversity for the RF  in Section III. We provide statistical results and performance analysis of the end-to-end system  with  MRC at the relay node in Section IV. Section V presents the computer simulations of the proposed work. Finally, Section VI summarizes  the main findings of the paper.

\emph{Notations}: $(\cdot)_i$ denotes the parameter of $i$-th RF link from the source to the relay, and  $(\cdot)$ denotes the parameter of the THz link from the relay to the destination. $\Gamma(a)= \int\limits_{0}^{\infty}t^{a-1} e^{-t}dt$  denotes the Gamma function.  $ \Gamma(a,t)=\int_{t}^{\infty}s^{a-1}e^{-s}ds$ and $ \gamma(a,t)=\int_{0}^{t}s^{a-1}e^{-s}ds$ are upper and lower incomplete Gamma functions, respectively. $ G_{p,q}^{m,n}\big(.|.\big)$ represents the Meijer's G-function while $ H_{p,q}^{m,n}\big(.|.\big) $ represents the Fox's H-function \cite{Mathai_2010}. We denote  $\{a_{1},\cdots,a_{N}\}$ by shorthand $\{a_{i}\}_{1}^{N} $. The imaginary number is denoted by $\J$.

\section{System Model}\label{sec:system_model}
\begin{figure}	
	\label{system_model}
	\includegraphics[width=\columnwidth]{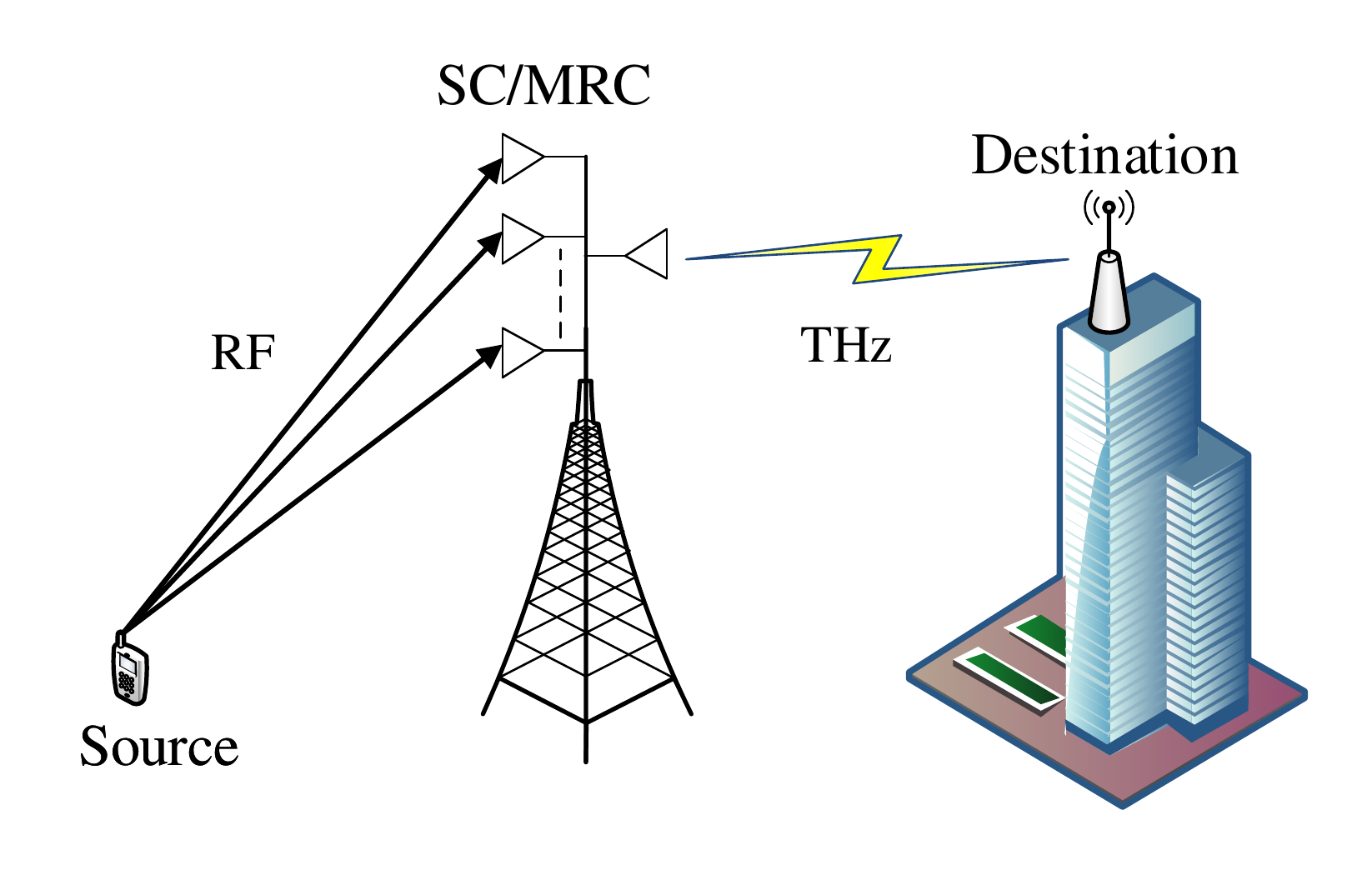}
		\vspace{-7mm}
	\caption{Hybrid multi-antenna RF and THz wireless transmission.}
\end{figure}
We consider a user in an access network that intends to transmit a signal  to a destination point in the backhaul link through an access point (AP) equipped with  multiple antennas ($N\geq 1$). The system model is readily applicable for the multi-user scenario (similar to the cell-free network \cite{Ngo_2017_cell_free} using  orthogonal resource allocation for different users. We use low frequency RF (typically $<10$ \mbox{GHz}) transmission from the source to the AP and THz band ($0.1$- $10$ \mbox{THz}) for signal relaying from the AP to the destination. 

We  use generalized  $\alpha$-$\mu$ distribution to model instantaneous SNR of  the RF link from  the source to the $i$-th antenna   \cite{Yacoub_alpha_mu}:
\begin{equation} \label{eq:pdf_hf_rf}
	f_{\gamma_{i}}(\gamma) = \frac{A_{i} \gamma^{\frac{\alpha_i\mu_i}{2}-1}}{ 2 {\bar{\gamma}_{\rm RF}}^{\frac{\alpha_i\mu_i}{2}} }  \exp \Big(- B_{i} {\Big(\frac{\gamma}{\bar{\gamma}_{\rm RF}}\Big)^{\frac{\alpha_i}{2}}}\Big)
\end{equation}
where $A_i = \frac{\alpha_{i}\mu_i^{\mu_{i}}}{\Omega^{\alpha_{i}\mu_{i}} \Gamma(\mu_{i})}$,  $B_i = \frac{\mu_{i}}{\Omega^{\alpha_{i}}}$,  $i=1,2,\cdots,N$, and  $\{\alpha_i, \mu_i, \Omega_i\}$ are the channel fading parameters  from the source to the $i$-th antenna of the RF  link, and $\bar{\gamma}_{\rm RF}$ is the average SNR of the RF link. The $\alpha$-$\mu$ distribution is a general fading model, which accounts for the non-linearity of a non-homogeneous propagation environment. The power parameter $\alpha$ represents the nonlinear characteristics of the fading and the parameter $\mu$ depicts  number of multipath clusters.
Using \eqref{eq:pdf_hf_rf}, the CDF of the SNR is given by
\begin{eqnarray}\label{eq:cdf_rf}  
	F_{\gamma_{i}}(\gamma) =  1-\Bigg(\frac{\Gamma\big(\mu_i, B_{i} \gamma^{\frac{\alpha_i}{2}}\big)}{\Gamma (\mu_i) {\bar{\gamma}_{\rm RF}}^{\frac{\alpha_i}{2}}}\Bigg)
\end{eqnarray}

To model the THz backhaul link, we consider i.ni.d $\alpha$-$\mu$ channel fading and path loss due to the atmospheric loss combined with a statistical model for antenna misalignment errors. The PDF of the combined effect of channel fading and pointing errors for the THz link is derived in \cite{Boulogeorgos_Analytical}\footnote{There is a typo in the equation (26) of \cite{Boulogeorgos_Analytical}. It should be $\gamma^2 A_0^{-\gamma^2} \frac{\mu^{\frac{\gamma^2}{\alpha}}}{\hat{h}_f^\phi \Gamma(\mu)} x^{\gamma^2-1} \Gamma\Big( \frac{\alpha\mu-\gamma^2}{\alpha}, \mu\frac{x^\alpha}{\hat{h}_f^\alpha} A_0^{-\alpha} \Big)$. }
\begin{eqnarray} \label{eq:pdf_hfp}
	f_{\gamma_{\rm THz}}(\gamma) =  \frac{A \phi S_0^{-\phi} \gamma^{\frac{\phi}{2}-1} }{2 {\bar{\gamma}_{\rm THz}}^{\frac{\phi}{2}}} \Gamma \bigg(\frac{\alpha \mu - \phi}{\alpha},  \frac{S_0^{-\alpha} \gamma^{\frac{\alpha}{2}}}{ {\bar{\gamma}_{\rm THz}}^{\frac{\alpha}{2}}}  \bigg)
\end{eqnarray}	
	where $\bar{\gamma}_{\rm THz}$ is the average SNR of the THz link, $A = \frac{\mu ^{\frac{\phi}{\alpha} }}{\Omega^{\phi } \Gamma(\mu )}$, and $B = \frac{\mu }{\Omega^{\alpha}}$ with $\{\alpha , \mu, \Omega\}$ are the channel fading parameters  from the AP to the destination. The term $S_0$ denotes the fraction of collected power and $\phi$ denotes the ration of normalized beam-width to jitter. The CDF of the THz link combined with pointing error for a general $\mu$ is given by \cite{Pranay_2021_TVT}
\begin{flalign}	\label{eq:cdf_hfp}
	F_{\gamma_{\rm THz}}(\gamma)&=  \frac{A   B^{-\frac{\phi}{\alpha}}  \phi}{\phi} \bigg[ \gamma\bigg(\mu,\frac{B} {S_0^{\alpha}}\Big(\sqrt{\frac{\gamma}{\bar{\gamma}_{\rm THz}}}\Big)^{\alpha}\bigg) \hspace{-1mm} \nonumber\\&+\frac{B^{\frac{\phi}{\alpha}}}{S_0^{\phi}}\Big(\sqrt{\frac{\gamma}{\bar{\gamma}_{\rm THz}}}\Big)^{\hspace{-1mm}\phi}\times \Gamma\Big(\frac{\alpha\mu-\phi}{\alpha},\frac{B} {S_0^{\alpha}}\Big(\sqrt{\frac{\gamma}{\bar{\gamma}_{\rm THz}}}\Big)^{\hspace{-1mm}\alpha}\Big) \bigg]
\end{flalign}

Finally,  we employ a frequency up-converter to match the carrier frequency of the RF with the THz and a  fixed gain AF relay to forward the signal from the source to the destination. The end-to-end SNR for the AF relaying system is given by 	$\gamma = \frac{\gamma_{\rm RF}\gamma_{\rm THz}}{\gamma_{\rm THz}+C}$ where $C$ can be obtained statistically from the received signal of the RF link and transmit power at the relay. Given the PDF of SNR  for individual links as $f_{\gamma_{\rm RF}}(\gamma)$ and $f_{\gamma_{\rm THz}}(\gamma)$, the PDF of the end-to-end SNR $\gamma$ for the dual-hop system can be represented as \cite{Zedini2016_fso}
\begin{flalign} \label{eq:pdf_eqn_af}
	&f_\gamma(z) = \int_{0}^{\infty}  {f_{\gamma_{\rm RF}}\left(\frac{z(x + C)}{x}\right)} {f_{\gamma_{\rm THz}}(x)} \frac{x + C}{{x}} {dx}
\end{flalign}
Similarly, the CDF of the end-to-end system using the CDF of individual links $F_{\gamma_{\rm RF}}(\gamma)$ and $F_{\gamma_{\rm THz}}(\gamma)$ is \cite{Zedini2016_fso}:
\begin{eqnarray}\label{eq:cdf_eqn_af}
	F_{\gamma}^{} (\gamma) = F_{\gamma_{\rm RF}} (\gamma) + \int_{\gamma}^{\infty} F_{\gamma_{\rm THz}} \big(\frac{C\gamma}{x-\gamma}\big)  f_{\gamma_{\rm RF}} (x) dx
\end{eqnarray} 
In the following sections, we derive the PDF of the RF link denoted as $f_{\gamma_{\rm RF}}(\gamma)$ using both SC and MRC diversity  techniques to develop  statistical results on the end-to-end SNR using the general representation in   \eqref{eq:pdf_eqn_af} and \eqref{eq:cdf_eqn_af}. 

\section{Performance of SC-RF and THz}
In this section, we employ selection combining diversity receiver for the multi-antenna RF and study the performance when mixed with the THz wireless link using the fixed-gain AF relaying. Denote by $\gamma_{i}$, where $i=1, 2, \cdots N$ the SNR between the source and the $i$-th antenna using RF transmission in the first hop. The selection combining opportunistically selects a single antenna with  maximum SNR to harness the spatial diversity. Thus, the SNR in the first link using the selection combining is given as $ \gamma_{\rm RF} = \max (\gamma_{1}, \gamma_{2}, ...\gamma_{N}) $. Assuming independent but non-identical fading, the CDF $F_{\gamma_{\rm RF}} (\gamma)$ of the end-to-end SNR of the first link is  given by  $F_{\gamma_{\rm RF}} (\gamma)  = \prod_{i=1}^NF_{\gamma_{i}} (\gamma)$, where $F_{\gamma_{i}} (\gamma)$ can be obtained from \eqref{eq:cdf_rf}. Assuming identical fading characteristics among multi-antenna system (which is a reasonable assumption and has been widely considered in the literature), the CDF of the first link can be expressed as
\begin{equation}\label{eq:cdf_max_eqn}
	F_{\gamma_{\rm RF}} (\gamma)  = [F_{\gamma_{i}} (\gamma)]^N
\end{equation}
Substituting \eqref{eq:cdf_max_eqn} in \eqref{eq:cdf_eqn_af}, we can develop the CDF of the mixed link using multivariate Fox's-H function.   In the following Theorem, we develop novel statistical results on the mixed link using bivariate Fox's H-function. 
\begin{my_theorem} \label{th:sc_af}
	The PDF and CDF of end-to-end SNR of the RF link with selection combining mixed with  THz using the  fixed-gain AF relay is given in \eqref{eq:pdf_snr_af} and \eqref{eq:cdf_snr_af}, respectively.
	\begin{figure*}
		\begin{flalign} \label{eq:pdf_snr_af}
			&f_{\gamma}^{\rm SC} (z) =  \frac{NAA_i\phi S_0^{-\alpha\mu } C^{\frac{\alpha\mu}{2}} }{ 2 \alpha^2 {{\bar{\gamma}_{\rm RF}}^{\frac{\alpha_i\mu_i}{2}}} {\bar{\gamma}_{\rm THz}}^{\frac{\alpha\mu}{2}}} \sum_{k_0+k_1+\cdots+k_{\mu_i-1}=N-1}^{}  \Big(\begin{matrix} N-1 \\ k_0+k_1+\cdots+k_{\mu_i-1}\end{matrix}\Big)   \prod_{t=o}^{\mu_i-1} \frac{B^{k_t} (z)^{(\sum_{t=0}^{\mu_i-1}\frac{t\alpha_i k_t}{2}+\frac{\alpha_i\mu_i}{2}-1)}}{(t!)^{k_t} {\bar{\gamma}_{\rm RF}}^{(\frac{t\alpha_ik_t}{2})}}  \nonumber \\ &  H_{1,0:1,1:2,4}^{0,1:0,1:3,1} \Bigg[ \begin{matrix}	\big\{\big(1-\frac{\alpha\mu}{2} + \sum_{t=0}^{\mu_i-1}\frac{t\alpha_i k_t}{2}+\frac{\alpha_i\mu_i}{2};\frac{\alpha_i}{2}, \frac{\alpha}{2}\big)\}; \big\{\big(1,1\big)\}; \big\{ \big(1-\mu, 1\big), \big(1-\mu+\frac{\phi}{\alpha}, 1\big) \}\\ \big\{-\} ; \big\{\big( 1+\sum_{t=0}^{\mu_i-1}\frac{t\alpha_i k_t}{2}+\frac{\alpha_i\mu_i}{2},\frac{\alpha_i}{2}\big)\} ; \big\{ \big(-\mu+\frac{\phi}{\alpha},1\big), \big(0,1\big),\big(-\frac{\alpha\mu}{2}, \frac{\alpha}{2}\big) \big(-\mu,1\big)\}\end{matrix}\Bigg|\frac{{{\bar{\gamma}_{\rm RF}}}^{\frac{\alpha_i}{2}}}{N z^{\frac{\alpha_i}{2}} B_i} ; \frac{ B C^{\frac{\alpha}{2}}}{S_o^{\alpha} {\bar{\gamma}_{\rm THz}}^{\frac{\alpha}{2}}} \Bigg] 
		\end{flalign}
	\end{figure*}
	
	\begin{figure*}
		\footnotesize
		\begin{flalign} \label{eq:cdf_snr_af} 
			&F_{\gamma}^{\rm SC} (\gamma) =  \frac{NAA_i\phi S_0^{-\alpha\mu } C^{\frac{\alpha\mu}{2}} }{ 2 \alpha^2 {{\bar{\gamma}_{\rm RF}}^{\frac{\alpha_i\mu_i}{2}}} {\bar{\gamma}_{\rm THz}}^{\frac{\alpha\mu}{2}}} \sum_{k_0+k_1+\cdots+k_{\mu_i-1}=N-1}^{}  \Big(\begin{matrix} N-1 \\ k_0+k_1+\cdots+k_{\mu_i-1}\end{matrix}\Big)   \prod_{t=o}^{\mu_i-1} \frac{B^{k_t} (\gamma)^{(\sum_{t=0}^{\mu_i-1}\frac{t\alpha_i k_t}{2}+\frac{\alpha_i\mu_i}{2})}}{(t!)^{k_t} {\bar{\gamma}_{\rm RF}}^{(\frac{t\alpha_ik_t}{2})}}  \nonumber \\ &  H_{1,0:2,2:2,4}^{0,1:1,1:3,1} \Bigg[ \begin{matrix}\big\{\big(1-\frac{\alpha\mu}{2} + \sum_{t=0}^{\mu_i-1}\frac{t\alpha_i k_t}{2}+\frac{\alpha_i\mu_i}{2};\frac{\alpha_i}{2}, \frac{\alpha}{2}\big)\}; \big\{\big(1,1\big), \big( 1+\sum_{t=0}^{\mu_i-1}\frac{t\alpha_i k_t}{2}+\frac{\alpha_i\mu_i}{2},\frac{\alpha_i}{2}\big)\}; \big\{ \big(1-\mu, 1\big), \big(1-\mu+\frac{\phi}{\alpha}, 1\big) \}\\ \big\{-\} ; \big\{\big(\sum_{t=0}^{\mu_i-1}\frac{t\alpha_i k_t}{2}+\frac{\alpha_i\mu_i}{2},\frac{\alpha_i}{2}\big), \big( 1+\sum_{t=0}^{\mu_i-1}\frac{t\alpha_i k_t}{2}+\frac{\alpha_i\mu_i}{2},\frac{\alpha_i}{2}\big) \} ; \big\{ \big(-\mu+\frac{\phi}{\alpha},1\big), \big(0,1\big),\big(-\frac{\alpha\mu}{2}, \frac{\alpha}{2}\big) \big(-\mu,1\big)\}\end{matrix}  \Bigg| \frac{{{\bar{\gamma}_{\rm RF}}}^{\frac{\alpha_i}{2}}}{N \gamma^{\frac{\alpha_i}{2}} B_i} ; \frac{ B C^{\frac{\alpha}{2}}}{S_o^{\alpha} {\bar{\gamma}_{\rm THz}}^{\frac{\alpha}{2}}} \Bigg] 
		\end{flalign}
		\normalsize
		\hrule
	\end{figure*}	
\end{my_theorem}

\begin{IEEEproof}
See Appendix A.
\end{IEEEproof}
Note that computation of the bivariate Fox's H-function is available in MATHEMATICA and MATLAB. In the following subsections, we use the statistical result of Theorem 1 to analyze the performance of the mixed RF-THz system.
\subsection{Outage Probability}
Outage probability is an important performance metric that describes the reliability of a communication system by comparing the received SNR with a threshold SNR $\gamma_{\rm th}$:  $ P_{\rm out} = P(\gamma <\gamma_{th})$. We can get an exact expression of the outage probability  by substituting  $\gamma = \gamma_{\rm th}$ in the CDF \eqref{eq:cdf_snr_af} in terms of bivariate Fox's H-function. We also simplify the expression of the  outage probability in high SNR regime by deriving $\lim_{\bar{\gamma} \to \infty}P_{\rm out}$, where $\bar{\gamma}=\bar{\gamma}_{\rm RF}=\bar{\gamma}_{\rm THz} $.  We apply \cite[Th. 1.7, 1.11]{Kilbas_2004} by calculating  residues at poles $S_1 = \frac{\alpha\mu- \sum_{t=0}^{\mu_i-1}t\alpha_i k_t-\alpha_i\mu_i+\alpha S_2 }{\alpha_i}, 0 $ and $S_2 = \frac{-\alpha\mu+ \sum_{t=0}^{\mu_i-1}t\alpha_i k_t+\alpha_i\mu_i }{\alpha}, -\mu+\frac{\phi}{\alpha}, 0, -\mu$ to get the asymptotic outage probability in \eqref{eq:outage_asymp_sc_af}. Applying  $\bar{\gamma}_{\rm RF}\to \infty$ and $\bar{\gamma}_{\rm THz} \to \infty $ in \eqref{eq:outage_asymp_sc_af},  the diversity order can be obtained as 
\begin{flalign}\label{eq:do_iid}
	DO_{}^{\rm SC} =& \biggl\{\frac{\alpha_i \mu_i}{2}, \frac{\phi-\sum_{t=0}^{\mu_i-1}t\alpha_i k_t}{2}, \frac{\alpha\mu- \sum_{t=0}^{\mu_i-1}t\alpha_i k_t }{2}, \nonumber \\ & \frac{-\sum_{t=0}^{\mu_i-1}t\alpha_i k_t}{2} \biggr\}
\end{flalign}
The diversity order in \eqref{eq:do_iid} is applicable for i.i.d fading scenario for multi-antenna RF transmissions with an  implicit presence of  spatial diversity parameter $N$. to To derive the diversity order for the general scenario, we can directly use  \eqref{eq:cdf_eqn_af} with asymptotic expansion of incomplete gamma function. We  use  \eqref{eq:cdf_rf} in $F_{\gamma_{\rm RF}} (\gamma)  = \prod_{i=1}^NF_{\gamma_{i}} (\gamma)$ and  apply the asymptotic expression of $\lim _{t \to 0}\gamma(a,t) = t^a/a$ \cite{Jameson2016} with the identity  $\Gamma(a,t)=\Gamma(a)-\gamma(a,t)$ to get the diversity order of the RF ink as $\frac{1}{2}{\sum_{i=1}^{N}\alpha_i\mu_i}$.  Substituting \eqref{eq:cdf_rf} and  \eqref{eq:cdf_hfp} in   $\int_{\gamma}^{\infty} F_{\gamma_{\rm THz}} \big(\frac{C\gamma}{t-\gamma}\big)  f_{\gamma_{\rm RF}} (t) dt$ with $\lim _{t \to \infty}\Gamma(a,t) = e^{-t}t^{a-1}$ and $\lim _{t \to 0}\gamma(a,t) = t^a/a$, the integral  can be simplified as $\min\bigl\{\frac{\alpha \mu}{2}, \frac{\phi}{2}\bigr\}$.  Thus, the diversity order of the RF-THz is given by  $\min\bigl\{\frac{\sum_{i=1}^{N}\alpha_i\mu_i}{2}, \frac{\alpha\mu}{2}, \frac{\phi}{2}\bigr\}$.  

The diversity order provides an important insight on the system deployment: the performance of the hybrid system can be made independent from RF fading and pointing errors by employing a sufficient number of antennas for RF reception and higher beam-width for THz transmissions.

\begin{figure*}
	\begin{flalign} \label{eq:outage_asymp_sc_af}
		P_{\rm out}^{{\rm SC}, \infty} &=  \frac{NAA_i\phi S_0^{-\alpha\mu } C^{\frac{\alpha\mu}{2}} }{ 2 \alpha^2 {{\bar{\gamma}}^{\frac{\alpha_i\mu_i}{2}}} {\bar{\gamma}}^{\frac{\alpha\mu}{2}}} \sum_{k_0+k_1+\cdots+k_{\mu_i-1}=N-1}^{}  \Big(\begin{matrix} N-1 \\ k_0+k_1+\cdots+k_{\mu_i-1}\end{matrix}\Big)   \prod_{t=o}^{\mu_i-1} \frac{B_i^{k_t} (\gamma)^{(\sum_{t=0}^{\mu_i-1}\frac{t\alpha_i k_t}{2}+\frac{\alpha_i\mu_i}{2})} }{(t!)^{k_t} {\bar{\gamma}}^{(\frac{t\alpha_ik_t}{2})}}     \nonumber \\ &  \times    \Bigg[ \Bigg( \frac{ \Gamma\big(\frac{ \sum_{t=0}^{\mu_i-1}t\alpha_i k_t+\alpha_i\mu_i  }{2}\big)  \Gamma\big(\frac{\alpha\mu- \sum_{t=0}^{\mu_i-1}t\alpha_i k_t-\alpha_i\mu_i }{\alpha}\big)} {\Gamma\big(\frac{2+ \sum_{t=0}^{\mu_i-1}t\alpha_i k_t+\alpha_i\mu_i }{2}\big) \big(\frac{\phi- \sum_{t=0}^{\mu_i-1}t\alpha_i k_t-\alpha_i\mu_i }{\alpha}\big) \big(\frac{ \sum_{t=0}^{\mu_i-1}t\alpha_i k_t+\alpha_i\mu_i }{\alpha}\big)}  \bigg( \frac{B C^{\frac{\alpha}{2}} }{S_o^{\alpha}{\bar{\gamma}}^{\frac{\alpha}{2}}}\bigg)^{\frac{-\alpha\mu+ \sum_{t=0}^{\mu_i-1}t\alpha_i k_t+\alpha_i\mu_i }{\alpha}} \nonumber \\ &  + \frac{\Gamma\big( \frac{- \sum_{t=0}^{\mu_i-1}t\alpha_i k_t-\alpha_i\mu_i+ \phi }{2}\big)  \Gamma\big(\sum_{t=0}^{\mu_i-1}\frac{t\alpha_i k_t+\alpha_i\mu_i}{2}\big)} {\Gamma\big(\frac{-\sum_{t=0}^{\mu_i-1}t\alpha_i k_t-\alpha_i\mu_i}{2}\big) \Gamma\big(\frac{2+\sum_{t=0}^{\mu_i-1}t\alpha_i k_t+\alpha_i\mu_i }{2}\big)} \frac{\Gamma\big(\mu-\frac{\phi}{\alpha}\big)  \Gamma\big(\frac{\phi}{2}\big)  }{ \big(\frac{\phi}{\alpha}\big)} \bigg( \frac{B C^{\frac{\alpha}{2}} }{S_o^{\alpha} {\bar{\gamma}}^{\frac{\alpha}{2}}}\bigg)^{-\mu+\frac{\phi}{\alpha}} \nonumber \\ &  +  \frac{\Gamma\big(\frac{\alpha\mu- \sum_{t=0}^{\mu_i-1}t\alpha_i k_t-\alpha_i\mu_i }{2}\big)  \Gamma\big(\sum_{t=0}^{\mu_i-1}\frac{t\alpha_i k_t+\alpha_i\mu_i}{2}\big)} {\Gamma\big(\frac{-\sum_{t=0}^{\mu_i-1}t\alpha_i k_t-\alpha_i\mu_i}{2}\big) \Gamma\big(\frac{2+\sum_{t=0}^{\mu_i-1}t\alpha_i k_t+\alpha_i\mu_i }{2}\big)}           \frac{  \Gamma\big(\frac{-\alpha\mu}{2}\big)}{(\mu+\frac{\phi}{\alpha}) \big(\mu\big)}   +        \frac{\Gamma\big(\sum_{t=0}^{\mu_i-1}\frac{t\alpha_ik_t+\alpha_i\mu_i}{2}\big)}{\Gamma\big(\frac{2+\sum_{t=0}^{\mu_i-1}t\alpha_ik_t+\alpha_i\mu_i}{2}\big)}\frac{\Gamma\big(\mu\big)}{(\frac{\phi}{\alpha})} \bigg( \frac{B C^{\frac{\alpha}{2}} }{S_o^{\alpha} {\bar{\gamma}}^{\frac{\alpha}{2}}}\bigg)^{-\mu}   \Bigg) \Big({{\bar{\gamma}}^{-\frac{\alpha_i\mu_i}{2}}}	\Big)   \nonumber \\ &   +    \Bigg( \frac{\Gamma\big(\frac{ \sum_{t=0}^{\mu_i-1}t\alpha_i k_t+\alpha_i\mu_i- {\phi} }{\alpha_i}\big) \Gamma\big(\frac{\phi}{2}\big)} {\Gamma\big(-\frac{\phi}{2}\big) \Gamma\big(\frac{2+\phi }{2}\big)}  \bigg(\frac{N \gamma^{\frac{\alpha_i}{2}} B_i}{ {{\bar{\gamma}}}^{\frac{\alpha_i}{2}}}\bigg)^{\frac{- \sum_{t=0}^{\mu_i-1}t\alpha_i k_t-\alpha_i\mu_i+\phi }{\alpha_i}} \frac{\Gamma\big(\frac{\phi}{\alpha}\big)\Gamma\big(\frac{\phi}{2}\big)}{ \big(\frac{\phi}{\alpha}\big)} \bigg(\frac{BC^{\frac{\alpha}{2}}}{S_o^{\alpha}{\bar{\gamma}}^{\frac{\alpha}{2}}}\bigg)^{-\mu+\frac{\phi}{\alpha}}\bigg)  \Big({\bar{\gamma}}^{\frac{\sum_{t=0}^{\mu_i-1}t\alpha_i k_t}{2}} {\bar{\gamma}}^{-\frac{\phi }{2}} \Big)  \nonumber \\ & + \Bigg( \frac{\Gamma\big(\frac{-\alpha\mu+ \sum_{t=0}^{\mu_i-1}t\alpha_ik_t+\alpha_i\mu_i}{\alpha_i}\big) \Gamma\big(\frac{\alpha\mu }{2}\big)} {\Gamma\big(\frac{-{\alpha\mu } }{2}\big) \Gamma\big(\frac{2+{\alpha\mu } }{2}\big)}\bigg(\frac{N\gamma^{\frac{\alpha_i}{2}}B_i}{{{\bar{\gamma}}}^{\frac{\alpha_i}{2}}}\bigg)^{\frac{\alpha\mu- \sum_{t=0}^{\mu_i-1}t\alpha_i k_t-\alpha_i\mu_i }{\alpha_i}} \frac{ \Gamma\big(\frac{-\alpha\mu}{2}\big)  }{(-\mu+\frac{\phi}{\alpha}) \big(\mu\big)}  \Bigg) \Big({\bar{\gamma}}^{-\frac{\alpha\mu}{2}}  {\bar{\gamma}}^{\frac{ \sum_{t=0}^{\mu_i-1}t\alpha_i k_t }{2}}\Big)  \nonumber \\ & +  \Bigg( {\Gamma\bigg(\frac{ \sum_{t=0}^{\mu_i-1}t\alpha_i k_t+\alpha_i\mu_i}{\alpha_i}\bigg)}  \bigg(\frac{N \gamma^{\frac{\alpha_i}{2}} B_i}{ {{\bar{\gamma}}}^{\frac{\alpha_i}{2}}}\bigg)^{\frac{- \sum_{t=0}^{\mu_i-1}t\alpha_i k_t-\alpha_i\mu_i}{\alpha_i}}   \frac{\Gamma\big(\mu\big)}{(\frac{\phi}{\alpha})} \bigg( \frac{B C^{\frac{\alpha}{2}} }{S_o^{\alpha} {\bar{\gamma}}^{\frac{\alpha}{2}}}\bigg)^{-\mu}  \Bigg) \Big({\bar{\gamma}}^{\frac{ \sum_{t=0}^{\mu_i-1}t\alpha_i k_t}{2}}\Big)    \Bigg]
	\end{flalign}
	\hrule
\end{figure*}
\subsection{Average BER}
The average BER of a communication system quantifies the detection error of the transmitted symbols: 
\begin{eqnarray}  \label{eq:ber_eqn}
	\bar{P}_e = \frac{q^p}{2\Gamma(p)}\int_{0}^{\infty} \gamma^{p-1} {e^{{-q \gamma}}} F_{\gamma} (\gamma)   d\gamma
\end{eqnarray}
where $p$ and $q$ determine the type of modulation. To derive the BER in a closed form, we substitute the CDF of \eqref{eq:cdf_snr_af} in \eqref{eq:ber_eqn} to get
\begin{flalign} 
	&\bar{P}_e^{\rm SC} =  \frac{NAA_i\phi S_0^{-\alpha\mu } C^{\frac{\alpha\mu}{2}} q^p}{ 4 \Gamma(p) \alpha^2 {{\bar{\gamma}_{\rm RF}}^{\frac{\alpha_i\mu_i}{2}}} {\bar{\gamma}_{\rm THz}}^{\frac{\alpha\mu}{2}}} \sum_{k_0+k_1+\cdots+k_{\mu_i-1}=N-1}^{} \nonumber \\ & \Big(\begin{matrix} N-1 \\ k_0+k_1+\cdots+k_{\mu_i-1}\end{matrix}\Big)   \prod_{t=o}^{\mu_i-1} \frac{B_i^{k_t} }{(t!)^{k_t} {\bar{\gamma}_{\rm RF}}^{(\frac{t\alpha_ik_t}{2})}}  \nonumber \\ &   \frac{\Gamma\big(-\frac{\alpha\mu}{2}-\frac{\alpha S_2}{2}\big)  \Gamma\big( \frac{\alpha\mu}{2} -\sum_{t=0}^{\mu_i-1}\frac{t\alpha_i k_t}{2}-\frac{\alpha_i\mu_i}{2}+\frac{\alpha_iS_1}{2} +\frac{\alpha S_2}{2}\big)}{\Gamma\big(-\sum_{t=0}^{\mu_i-1}\frac{t\alpha_ik_t}{2}-\frac{\alpha_i\mu_i}{2}+\frac{\alpha_iS_1}{2}\big)} \nonumber \\ &  \frac{ \Gamma(\sum_{t=0}^{\mu_i-1}\frac{t\alpha_i k_t}{2}+\frac{\alpha_i\mu_i}{2}-\frac{\alpha_iS_1}{2}) }{\Gamma(1+\sum_{t=0}^{\mu_i-1}\frac{t\alpha_i k_t}{2}+\frac{\alpha_i\mu_i}{2}-\frac{\alpha_iS_1}{2})} \nonumber \\ & \frac{1}{2\pi \J} \int_{\mathcal{L}_1} \Gamma(S_1) \bigg(\frac{ {{\bar{\gamma}_{\rm RF}}}^{\frac{\alpha_i}{2}}}{NB_i}\bigg)^{S_1} dS_1 \nonumber \\ &  \frac{1}{2\pi \J} \hspace{-1mm} \int_{\mathcal{L}_2} \hspace{-1mm}\frac{\Gamma(-\mu+\frac{\phi}{\alpha}-S_2) \Gamma(-S_2) \Gamma(\mu+S_2) }{\Gamma(1-\mu+\frac{\phi}{\alpha}-S_2) \Gamma(1+\mu+S_2)} \bigg( \frac{B C^{\frac{\alpha}{2}} }{S_o^{\alpha} {\bar{\gamma}_{\rm THz}}^{\frac{\alpha}{2}}}\bigg)^{S_2} \hspace{-1mm} dS_2 I_2
\end{flalign}
The inner integral $I_2 = \int_{0}^{\infty} {e^{{-q \gamma}}} \gamma^{(\sum_{t=0}^{\mu_i-1}\frac{t\alpha_i k_t}{2}+\frac{\alpha_i\mu_i}{2}-\frac{\alpha_iS_1}{2}+p-1)}d\gamma$ can be solved to $q^{-(\sum_{t=0}^{\mu_i-1}\frac{t\alpha_i k_t}{2}+\frac{\alpha_i\mu_i}{2}-\frac{\alpha_iS_1}{2})-p} \Gamma\big(\sum_{t=0}^{\mu_i-1}\frac{t\alpha_i k_t}{2}+\frac{\alpha_i\mu_i}{2}-\frac{\alpha_iS_1}{2}+p\big)$ using \cite[3.381/4]{Gradshteyn} in terms of Gamma function. Finally, applying the definition of Fox's H-function we get the average BER in \eqref{eq:ber_sc_af}.  
\begin{flalign} \label{eq:ber_sc_af}
	\bar{P}_e^{\rm SC} &=  \frac{NAA_i\phi S_0^{-\alpha\mu } C^{\frac{\alpha\mu}{2}} }{ 4 \Gamma(p) \alpha^2 {{\bar{\gamma}_{\rm RF}}^{\frac{\alpha_i\mu_i}{2}}} {\bar{\gamma}_{\rm THz}}^{\frac{\alpha\mu}{2}}} \sum_{k_0+k_1+\cdots+k_{\mu_i-1}=N-1}^{} \nonumber \\ & \times \Big(\begin{matrix} N-1 \\ k_0+k_1+\cdots+k_{\mu_i-1}\end{matrix}\Big)   \prod_{t=o}^{\mu_i-1} \frac{B_i^{k_t} q^{-(\sum_{t=0}^{\mu_i-1}\frac{t\alpha_i k_t}{2}+\frac{\alpha_i\mu_i}{2})}}{(t!)^{k_t} {\bar{\gamma}_{\rm RF}}^{(\frac{t\alpha_ik_t}{2})}}  \nonumber \\ & \times H_{1,0:2,3:2,4}^{0,1:2,1:3,1} \Bigg[ \begin{matrix} ~~U_1~~ \\ ~~V_{1}~~ \end{matrix} \Bigg|\frac{ {{\bar{\gamma}_{\rm RF}}}^{\frac{\alpha_i}{2}} q^{\frac{\alpha_i}{2}}}{NB_i} ; \frac{ B C^{\frac{\alpha}{2}}}{S_o^{\alpha} {\bar{\gamma}_{\rm THz}}^{\frac{\alpha}{2}}} \Bigg] 
\end{flalign}
where $U_1 = \big\{\big(1-\frac{\alpha\mu}{2} + \sum_{t=0}^{\mu_i-1}\frac{t\alpha_i k_t}{2}+\frac{\alpha_i\mu_i}{2};\frac{\alpha_i}{2}, \frac{\alpha}{2}\big)\}; \big\{\big(1,1\big), \big( 1+\sum_{t=0}^{\mu_i-1}\frac{t\alpha_i k_t}{2}+\frac{\alpha_i\mu_i}{2},\frac{\alpha_i}{2}\big)\}; \big\{ \big(1-\mu, 1\big), \big(1-\mu+\frac{\phi}{\alpha}, 1\big) \} $ and $V_1 = \big\{-\} ; \big\{\big(\sum_{t=0}^{\mu_i-1}\frac{t\alpha_i k_t}{2}+\frac{\alpha_i\mu_i}{2}+p, \frac{\alpha_i}{2}\big),\big(\sum_{t=0}^{\mu_i-1}\frac{t\alpha_i k_t}{2}+\frac{\alpha_i\mu_i}{2},\frac{\alpha_i}{2}\big), \big( 1+\sum_{t=0}^{\mu_i-1}\frac{t\alpha_i k_t}{2}+\frac{\alpha_i\mu_i}{2},\frac{\alpha_i}{2}\big) \} ; \big\{ \big(-\mu+\frac{\phi}{\alpha},1\big), \big(0,1\big),\big(-\frac{\alpha\mu}{2}, \frac{\alpha}{2}\big) \big(-\mu,1\big)\} $.

The asymptotic average BER expression in the high SNR region can be obtained by following similar steps as adopted for the derivation of the  asymptotic expression for the  outage probability. 
\subsection{Ergodic Capacity}
The ergodic capacity can determine the average throughput of the system over fading channels:
\begin{eqnarray} \label{eq:capacity_eqn}
	\bar{\eta} =\int_{0}^{\infty}{\rm log_2}(1+\gamma)f_\gamma(\gamma) d\gamma 
\end{eqnarray}
We substitute the PDF of \eqref{eq:pdf_snr_af} in \eqref{eq:capacity_eqn} and use the integral representation of Fox's H-function with the inner integral as \cite[4.293.3]{Gradshteyn} 
\begin{flalign}
	&\int_{0}^{\infty} {\rm log_2}(1+\gamma) (\gamma)^{(\sum_{t=0}^{\mu_i-1}\frac{t\alpha_i k_t}{2}+\frac{\alpha_i\mu_i}{2}-\frac{\alpha_iS_1}{2}-1)} d\gamma \nonumber \\ &  = \frac{\pi Csc\big(\pi (\sum_{t=0}^{\mu_i-1}\frac{t\alpha_i k_t}{2}+\frac{\alpha_i\mu_i}{2}-\frac{\alpha_iS_1}{2}) \big)}{{\rm log}(2)(\sum_{t=0}^{\mu_i-1}\frac{t\alpha_i k_t}{2}+\frac{\alpha_i\mu_i}{2}-\frac{\alpha_iS_1}{2})}
\end{flalign}

Further, using the relation $\pi Csc(\pi a) = \Gamma(a)\Gamma(1-a)$, we denote the denominator in terms of compatible Gamma functions, and applying the definition of Fox's H-function, we get the ergodic capacity in \eqref{eq:capacity_sc_af}.
\begin{flalign} \label{eq:capacity_sc_af}
	\bar{\eta}^{\rm SC} = & \frac{NAA_i\phi S_0^{-\alpha\mu } C^{\frac{\alpha\mu}{2}} }{ 2 {\rm log}(2) \alpha^2 {{\bar{\gamma}_{\rm RF}}^{\frac{\alpha_i\mu_i}{2}}} {\bar{\gamma}_{\rm THz}}^{\frac{\alpha\mu}{2}}} \sum_{k_0+k_1+\cdots+k_{\mu_i-1}=N-1}^{} \nonumber \\ & \times \Big(\begin{matrix} N-1 \\ k_0+k_1+\cdots+k_{\mu_i-1}\end{matrix}\Big)   \prod_{t=o}^{\mu_i-1} \frac{B_i^{k_t}}{(t!)^{k_t} {\bar{\gamma}_{\rm RF}}^{(\frac{t\alpha_ik_t}{2})}}  \nonumber \\ & \times H_{1,0:3,3:2,4}^{0,1:2,2:3,1} \Bigg[ \begin{matrix} ~~U_2~~ \\ ~~V_2~~ \end{matrix} \Bigg| \frac{ {{\bar{\gamma}_{\rm RF}}}^{\frac{\alpha_i}{2}}}{NB_i} ; \frac{ B C^{\frac{\alpha}{2}}}{S_o^{\alpha} {\bar{\gamma}_{\rm THz}}^{\frac{\alpha}{2}}} \Bigg] 
\end{flalign}
where $U_2 = \big\{\big(1-\frac{\alpha\mu}{2} + \sum_{t=0}^{\mu_i-1}\frac{t\alpha_i k_t}{2}+\frac{\alpha_i\mu_i}{2};\frac{\alpha_i}{2}, \frac{\alpha}{2}\big)\}; \big\{\big(1,1\big), \big(\sum_{t=0}^{\mu_i-1}\frac{t\alpha_i k_t}{2}+\frac{\alpha_i\mu_i}{2},\frac{\alpha_i}{2}\big), \big(1+\sum_{t=0}^{\mu_i-1}\frac{t\alpha_i k_t}{2}+\frac{\alpha_i\mu_i}{2},\frac{\alpha_i}{2}\big)\}; \big\{ \big(1-\mu, 1\big), \big(1-\mu+\frac{\phi}{\alpha}, 1\big) \} $ and $V_2 = \big\{-\} ; \big\{ \big( \sum_{t=0}^{\mu_i-1}\frac{t\alpha_i k_t}{2}+\frac{\alpha_i\mu_i}{2},\frac{\alpha_i}{2}\big), \big( \sum_{t=0}^{\mu_i-1}\frac{t\alpha_i k_t}{2}+\frac{\alpha_i\mu_i}{2},\frac{\alpha_i}{2}\big) ,\big( 1+\sum_{t=0}^{\mu_i-1}\frac{t\alpha_i k_t}{2}+\frac{\alpha_i\mu_i}{2},\frac{\alpha_i}{2}\big)\} ; \big\{ \big(-\mu+\frac{\phi}{\alpha},1\big), \big(0,1\big),\big(-\frac{\alpha\mu}{2}, \frac{\alpha}{2}\big) \big(-\mu,1\big)\} $.

\begin{figure*}[t]
	\begin{flalign} \label{eq:outage_asymptotic}
	F_{\gamma}^{{\rm MRC}, \infty}& = \frac{AA_m \phi S_0^{-\alpha \mu } {\rm \gamma_{}}^{\frac{\alpha \mu }{2}}}{4\alpha } \Bigg[\Bigg(\frac{2 \Gamma\big(\frac{\phi-\alpha_m\mu_m}{2}\big) { \Gamma(\mu -\frac{\phi}{\alpha }) \Gamma(-\frac{\phi}{2})} C^{\frac{\phi}{2}}} {{\alpha_m\mu_m} \Gamma(\frac{-\alpha_m\mu_m}{2}) {\bar{\gamma}}^{\frac{\phi}{2}}} \Big(\frac{B}{{S_0}^{\alpha } } \Big)^{\frac{\phi}{\alpha }-\mu }  +  \frac{ 2 \Gamma\big(\frac{\alpha \mu -\alpha_m\mu_m}{2}\big) \Gamma(\frac{\phi}{\alpha }-\mu)  \Gamma(\frac{-\alpha \mu }{2}) C^{\frac{\alpha \mu }{2}} {\bar{\gamma}}^{-\frac{\alpha \mu }{2}} } {\Gamma(1+\frac{\phi}{\alpha }) {\alpha_m\mu_m} \Gamma(\frac{-\alpha_m\mu_m}{2})}  \nonumber \\ &  +  \frac{4 \Gamma(\mu )} { \phi \alpha_m\mu_m}
	\Big(\frac{B}{{S_0}^{\alpha }} \Big)^{-\mu }  + \frac{8\Gamma\big(\frac{\alpha \mu -\alpha_m\mu_m}{\alpha }\big) C^{\frac{\alpha_m\mu_m}{2}}} { \alpha_m (\phi-\alpha_m\mu_m) \alpha_m\mu_m {\bar{\gamma}}^{\frac{\alpha_m\mu_m}{2}}} \Big(\frac{B}{{S_0}^{\alpha }} \Big)^{\frac{\alpha_m\mu_m-\alpha \mu }{\alpha }} \Bigg) {\bar{\gamma}}^{-\frac{\alpha_m\mu_m}{2}} \nonumber \\ & + \Bigg(\frac{4 \Gamma\big(\mu -\frac{\phi}{\alpha }\big) \Gamma\big(\frac{-\phi+\alpha_m\mu_m}{\alpha_m}\big) {B_m}^{\frac{\phi}{\alpha_m}-\mu_m} C^{\frac{\phi}{2}}} {\alpha_m\phi {\bar{\gamma}}^\frac{\phi}{2}} \Big(\frac{B }{{S_0}^{\alpha }} \Big)^{\frac{\phi}{\alpha }-\mu }\Bigg) \big({\bar{\gamma}}\big)^{-\frac{\phi}{2}}  +  \Bigg(\frac{4 \Gamma\big(\frac{-\alpha \mu +\alpha_m\mu_m}{\alpha_m}\big)  {B_m}^{\frac{\alpha \mu -\alpha_m\mu_m}{\alpha_m}} C^{\frac{\alpha \mu }{2}}} {\alpha_m (\frac{\phi}{\alpha }-\mu ) \alpha \mu  {\bar{\gamma}}^{\frac{\alpha \mu }{2}} }\Bigg) {\bar{\gamma}}^{-\frac{\alpha \mu }{2}} \Bigg]
	\end{flalign} 
	\hrule
\end{figure*}

\section{Performance of MRC-RF and THz}
In this section, we analyze the performance of the hybrid system by employing the optimal MRC diversity technique,  which combines coherently the received signals at $N$-antennas. Thus, $\gamma_{\rm RF} = \sum_{i=1}^{N} {\gamma_{i}}$ is the resultant SNR of the RF link after the MRC. It should be noted that  performance analysis can also be carried out  when  equal-gain combining (EGC) based multi-antenna  diversity technique is applied  for the RF using the SNR $\gamma_{\rm RF} = \frac{1}{N} \Big(\sum_{i=1}^{N} \sqrt{\gamma_{i}}\Big)^2$. In the following theorem, we derive the exact PDF and CDF of the SNR for the hybrid system consisting of  the MRC-RF and THz:
\begin{my_theorem} \label{th:mrc_af}
The exact PDF and CDF of end-to-end SNR of the RF link with the MRC mixed with THz using the fixed-gain AF relay in terms of $N+1$ variate Fox's H-function is given by:	
\begin{eqnarray} \label{eq:pdf_af_MRC}
	&f_{\gamma}^{\rm MRC}(\gamma) = \prod_{i=1}^{N} \frac{A_{i} \gamma^{\big(\sum_{i=1}^{N}\frac{\alpha_i\mu_i}{2}-1\big)}  A\phi S_0^{-\alpha\mu} C^{\frac{\alpha\mu}{2}} }{2 {\bar{\gamma}_{\rm RF}}^{\frac{\alpha_i\mu_i}{2}}  \alpha^2 {\bar{\gamma}_{\rm THz}}^{\frac{\alpha\mu}{2}}}  \nonumber \\ &  H_{2,1;1,1;\cdots;1,1;2,4}^{0,1;1,1;\cdots;1,1;3,1} \left[\begin{matrix} \biggl\{\frac{{\bar{\gamma}_{\rm rf}}^{\frac{\alpha_i}{2}}} {B_i \gamma^{\frac{\alpha_i}{2}}}\biggr\}_{i=1}^{N} \\ \frac{B C^{\frac{\alpha}{2}}}{S_o^{\alpha} {\bar{\gamma}_{\rm THz}}^{\frac{\alpha}{2}}}   \end{matrix} \Bigg|  \begin{matrix}	~~U_3~~  \\  ~~V_3~~ 	\end{matrix}  \right]
\end{eqnarray}
where $ U_3 = \Bigl\{\Big(1-\frac{\alpha\mu}{2}+\sum_{i=1}^{N}\frac{\alpha_i\mu_i}{2}; \frac{\alpha_1}{2},\cdots,\frac{\alpha_i}{2},\frac{\alpha}{2} \Big); \Big(\sum_{i=1}^{N}\frac{\alpha_i\mu_i}{2}; \frac{\alpha_1}{2},\cdots,\frac{\alpha_i}{2},0 \Big)\Bigr\}: \bigl\{(1,1),\cdots,(1,1)_N \bigr\}; \bigl\{\big(1-\mu,1\big),\big(1-\mu=\frac{\phi}{\alpha},1\big)\bigr\}$, and $V_3 = \Bigl\{\Big(1+\sum_{i=1}^{N}\frac{\alpha_i\mu_i}{2}; \frac{\alpha_1}{2},\cdots,\frac{\alpha_i}{2},0 \Big)\Bigr\}:  \Bigl\{\Big(\frac{\alpha_{1}\mu_{1}}{2}, \frac{\alpha_{1}}{2}\Big),\cdots, \Big(\frac{\alpha_i\mu_i}{2}, \frac{\alpha_i}{2}\Big)\Bigr\}; \Bigl\{\big(-\mu+\frac{\phi}{\alpha},1\big), \big(0,1\big), \big(-\mu,1\big),\big(\frac{-\alpha\mu}{2}, \frac{\alpha}{2}\big)\Bigr\}$.

\begin{eqnarray} \label{eq:cdf_af_MRC}
	&F_{\gamma}^{\rm MRC}(\gamma) = \prod_{i=1}^{N} \frac{A_{i} \gamma^{\big(\sum_{i=1}^{N}\frac{\alpha_i\mu_i}{2}\big)} A\phi S_0^{-\alpha\mu } C^{\frac{\alpha\mu}{2}}}{2 {\bar{\gamma}_{\rm RF}}^{\frac{\alpha_i\mu_i}{2}}  \alpha^2 {\bar{\gamma}_{\rm THz}}^{\frac{\alpha\mu }{2}}}  \nonumber \\ &  H_{3,1;1,1;\cdots;1,1;2,4}^{0,1;1,1;\cdots;1,1;3,1}  \left[\begin{matrix} \biggl\{\frac{{\bar{\gamma}_{\rm RF}}^{\frac{\alpha_i}{2}}} {B_i \gamma^{\frac{\alpha_i}{2}}}\biggr\}_{i=1}^{N} \\ \frac{B C^{\frac{\alpha}{2}}}{S_o^{\alpha} {\bar{\gamma}_{\rm THz}}^{\frac{\alpha}{2}}}   \end{matrix} \Bigg|  \begin{matrix} ~~U_4~~  \\ ~~V_4~~ \end{matrix}  \right]
\end{eqnarray}
where  $U_4 = \Bigl\{\Big(1-\frac{\alpha\mu }{2}+\frac{\sum_{i=1}^{N}\alpha_i\mu_i}{2}; \frac{\alpha_1}{2},\cdots,\frac{\alpha_i}{2},\frac{\alpha}{2} \Big); \Big(\frac{\sum_{i=1}^{N}\alpha_i\mu_i}{2}; \frac{\alpha_1}{2},\cdots,\frac{\alpha_i}{2},0 \Big); \\ \Big(\frac{\sum_{i=1}^{N}\alpha_i\mu_i}{2}; \frac{\alpha_1}{2},\cdots,\frac{\alpha_i}{2},0 \Big)\Bigr\}: \bigl\{(1,1),\cdots,(1,1)_N\bigr\} ; \bigl\{\big(1-\mu,1\big),\big(1-\mu-\frac{\phi}{\alpha},1\big)\bigr\}$ and $V_4 = \Bigl\{\Big(1+\sum_{i=1}^{N}\frac{\alpha_i\mu_i}{2}; \frac{\alpha_{1}}{2},\cdots,\frac{\alpha_i}{2},0 \Big)\Bigr\}:  \Bigl\{\Big(\frac{\alpha_{1}\mu_{1}}{2}, \frac{\alpha_{1}}{2}\Big),\cdots, \Big(\frac{\alpha_i\mu_i}{2}, \frac{\alpha_i}{2}\Big)\Bigr\}; \Bigl\{\big(-\mu+\frac{\phi}{\alpha},1\big), \big(0,1\big), \big(-\mu,1\big),\big(\frac{-\alpha\mu }{2}, \frac{\alpha}{2}\big)\Bigr\}$.
\end{my_theorem}

\begin{IEEEproof}
	See Appendix B.
\end{IEEEproof}

The diversity order for the outage probability can be computed using the residue of the multivariate Fox's H-function  at  dominant poles with a specific representation \cite{Kilbas_2004}, \cite{AboRahama_2018}. However, the multivariate Fox's H-function in \eqref{eq:cdf_af_MRC} consists of $N$ variables with average SNR $\bar{\gamma}_{\rm RF}$  in the numerator, while the $N+1$-th variable contains the average SNR $\bar{\gamma}_{\rm THz}$ in the denominator, precluding the direct application of \cite{AboRahama_2018} for the asymptotic analysis.  In the following lemma,  we present the diversity order of the considered hybrid system:
\begin{my_lemma}
	The diversity order of the RF link with the MRC receiver mixed with  the THz over i.ni.d channel fading is:
	\begin{flalign} \label{eq:diversity_MRC_mv}
	DO^{\rm MRC}_{} = \min\biggl\{\frac{\sum_{i=1}^{N}\alpha_i\mu_i}{2}, \frac{\alpha \mu}{2}, \frac{\phi}{2}\biggr\}
	\end{flalign}
\end{my_lemma}

\begin{IEEEproof}
We decompose the analysis in two parts:  applying $\bar{\gamma}_{\rm THz}\to \infty $ in \eqref{eq:cdf_af_MRC} for the first $N$-variables to get dominant poles as $p_i = \frac{\alpha_i\mu_i}{2}$, $i=1, 2, \cdots, N$ and $\bar{\gamma}_{\rm RF}\to 0 $  in \eqref{eq:cdf_af_MRC} to  get $N+1$-th pole as $p_{N+1} = \min\bigl\{\frac{\alpha \mu}{2},\frac{\phi}{2}\}$. Computing residues of \eqref{eq:cdf_af_MRC} at these poles, we get
\begin{flalign}\label{eq:asym_nplus1}
	&P_{\rm out}^{{\rm MRC}, \infty}= \prod_{i=1}^{N} \frac{2A_{i} A\phi C^{\frac{\alpha\mu}{2}}}{\alpha_{i}^2  \alpha^3 B_{i}^{\mu_{i}} B^{\mu}}\nonumber\\& \frac{\Gamma(p_{1}+\cdots+p_{N})}{\Gamma(p_{1}+\cdots+p_{N})\Gamma(-p_{1}-\cdots-p_{N})\Gamma(-p_{1}-\cdots-p_{N})} \nonumber \\ &\Gamma\Big(1+\mu_{i}+\frac{2}{\alpha_{i}}p_{i}\Big) \bigg(\frac{{\bar{\gamma}_{\rm RF}}^{}} {B_i^{\frac{2}{\alpha_i}} \gamma_{\rm th}}\bigg)^{-p_{i}} \nonumber \\ &\frac{\Gamma(\frac{\phi}{\alpha}-\frac{2}{\alpha}p_{N+1}) \Gamma(\mu-\frac{2}{\alpha}p_{N+1}) \Gamma(-\frac{2}{\alpha}p_{N+1}) \Gamma(\frac{2}{\alpha}p_{N+1})}{\Gamma(1-\frac{\phi}{\alpha}-\frac{2}{\alpha}p_{N+1})\Gamma(1+p_{N+1})} \nonumber \\ & \bigg(\frac{B^{\frac{2}{\alpha}} C}{S_0^{2} {\bar{\gamma}_{\rm THz}}^{}}\bigg)^{p_{N+1}} 
\end{flalign}
Using  $\bar{\gamma}=\bar{\gamma}_{\rm RF}=\bar{\gamma}_{\rm THz} $ and combining the exponents of $\bar{\gamma}$ in \eqref{eq:asym_nplus1}, we  get the diversity order of the system, as given in \eqref{eq:diversity_MRC_mv}.
\end{IEEEproof}


Further, we can use the statistical results of Theorem \eqref{th:mrc_af} to derive the average BER and ergodic capacity of the hybrid system, however,  in terms of $N+1$ variate Fox's H-function. It is desirable to simplify the analysis and reduce the dimension of the Fox's H-function. The authors in \cite{Costa_2008_alpha_mu_sum} have shown that the sum of $\alpha$-$\mu$ variates can be accurately approximated by a single $\alpha$-$\mu$ distribution \cite{Moraes2008_alpha_mu_sum,Ayadi2014_alpha_mu_sum,Wang2015_alpha_mu_sum,Kong2018_alpha_mu_sum}. Hence, we use \cite{Costa_2008_alpha_mu_sum} to simplify the statistical performance of the hybrid RF-THz system. 

We use  \eqref{eq:pdf_hf_rf} in $\gamma_{\rm RF} = \sum_{i=1}^{N} {\gamma_{i}}$ to get the PDF of the SNR for the first link as
\begin{equation} \label{eq:pdf_hf_rf_mrc}
	f_{\gamma_{\rm RF}}(\gamma) \approx  \frac{A_m \gamma ^{\frac{\alpha_m\mu_m}{2}-1}} {2 {\bar{\gamma}_{\rm RF}}^{\frac{\alpha_m\mu_m}{2}}} \exp \bigg(- B_m {\Big(\sqrt{\frac{\gamma}{\bar{\gamma}}_{\rm RF}}\Big)^{\alpha_m}}\bigg)
\end{equation}
where $A_m = \frac{\alpha_{m}\mu_m^{\mu_{m}}}{\Omega^{\alpha_{m}\mu_{m}} \Gamma(\mu_{m})}$,  $B_m = \frac{\mu_{m}}{\Omega^{\alpha_{m}}}$,  and  $\{\alpha_m, \mu_m, \Omega_m\}$ can be obtained by moment matching method \cite{Costa_2008_alpha_mu_sum}. Using Meijer's G-representation of exponential function, we can express the PDF of SNR of the RF link $f_{\gamma_{\rm RF}}(\gamma)$ in \eqref{eq:pdf_hf_rf_mrc} as
\begin{eqnarray} \label{eq:pdf_hf_rf_MG}
	f_{\gamma_{\rm RF}}(\gamma) \approx \frac{A_m \gamma ^{\frac{\alpha_m\mu_m}{2}-1}} {2 {\bar{\gamma}_{\rm RF}}^{\frac{\alpha_m\mu_m}{2}}}  G_{0, 1}^{1,0}  \Bigg( \frac{B_m {\gamma }^{\frac{\alpha_m}{2}}} {{ {\bar{\gamma}_{\rm RF}}}^{\frac{\alpha_m}{2}}} \Bigg| \begin{matrix} -\\0\end{matrix}\Bigg)                          
\end{eqnarray}

Next, we substitute \eqref{eq:pdf_hfp} and \eqref{eq:pdf_hf_rf_MG} in \eqref{eq:pdf_eqn_af}, and  adopt  the similar procedure used in deriving Theorem \ref{th:sc_af} to get the PDF as
\begin{flalign}\label{pdf_relay_fox}
	f^{\rm MRC}_{\gamma}(\gamma) \approx &  \frac{A A_m \phi S_0^{-\alpha\mu }  {\bar{\gamma}_{\rm THz}}^{-{\frac{\alpha\mu }{2}}} C^{\frac{\alpha\mu }{2}} {\bar{\gamma}_{\rm RF}}^{-\frac{\alpha_m\mu_m}{2}} {\gamma}^{\frac{\alpha_m\mu_m-2}{2}}} {4\alpha} \nonumber \\ & H_{1,0:1,3:1,1}^{0,1:3,0:0,1} \left[ \frac{B {C}^{\frac{\alpha}{2}}}{S_0^{\alpha} {\bar{\gamma}_{\rm THz}}^{\frac{\alpha}{2}}}, \frac{ {\bar{\gamma}_{\rm RF}}^{\frac{\alpha_m}{2}}} {B_m \gamma^{\frac{\alpha_m}{2}}}\Bigg|\begin{matrix}~~~U_5~~~\\~~~V_5~~~\end{matrix}\right]
\end{flalign}
where $U_5 = \big\{\big(1-\big(\frac{\alpha\mu -\alpha_m\mu_m}{2}\big);\frac{\alpha}{2},\frac{\alpha_m}{2} \big)\big\} : \big\{ \big(1+\frac{\phi}{\alpha}-\mu ,1 \big)\big\} : \big\{\big(1,1\big) \big\}$, and $ V_5 = \big\{ - \big\} : \big\{ \big(\frac{\phi}{\alpha}-\mu ,1 \big),  \big(0,1\big), \big(\frac{-\alpha\mu }{2},\frac{\alpha}{2}\big)\big\}:\big\{\big(1+\frac{\alpha_m\mu_m}{2}, \frac{\alpha_m}{2}\big)\big\}$.

To derive an approximate expression of the outage probability for the RF-MRC and THz, we substitute \eqref{pdf_relay_fox} in $F_{\gamma}(\gamma_{}) = \int_{0}^{\gamma_{\rm th}} f_{\gamma}(z) {dz}$, and apply the definition of Fox's H function with
\begin{flalign}
	\int_{0}^{\gamma_{\rm th}} z^{\frac{\alpha_m\mu_m-\alpha_ms_2-2}{2}} dz=\frac {\gamma_{\rm th}^{{\frac{\alpha_m\mu_m-\alpha_ms_2}{2}}} \Gamma\big(\frac{\alpha_m\mu_m+\alpha_ms_2}{2}\big)}{\Gamma\big(1+\frac{\alpha_m\mu_m+\alpha_ms_2}{2}\big)}
\end{flalign}
to get the outage probability using the bivariate Fox's-H function: 
\begin{flalign}\label{cdf_relay_fox}
	P_{\rm out}^{\rm MRC} \approx  & \frac{A A_m \phi S_0^{-\alpha\mu }  {\bar{\gamma}_{\rm THz}}^{-{\frac{\alpha\mu }{2}}} C^{\frac{\alpha\mu }{2}} {\bar{\gamma}_{\rm RF}}^{-\frac{\alpha_m\mu_m}{2}} {\gamma_{\rm th}}^{\frac{\alpha_m\mu_m}{2}}} {4\alpha} \nonumber \\ & H_{1,0:1,3:2,2}^{0,1:3,0:1,1} \bigg[ \frac{B {C}^{\frac{\alpha}{2}}}{S_0^{\alpha} {\bar{\gamma}_{\rm THz}}^{\frac{\alpha}{2}}}, \frac{ {\bar{\gamma}_{\rm RF}}^{\frac{\alpha_m}{2}}} {B_m {\gamma_{\rm th}}^{\frac{\alpha_m}{2}}}  \bigg| \begin{matrix} ~~U_6~~ \\  ~~V_6~~ \end{matrix} \bigg]
\end{flalign}
where $U_6 = \big\{\big(1-\frac{\alpha\mu -\alpha_m\mu_m}{2};\frac{\alpha}{2},\frac{\alpha_m}{2} \big)\big\} : \big\{ \big(1+\frac{\phi}{\alpha}-\mu ,1 \big)\big\} : \big\{\big(1,1\big), \big(1+\frac{\alpha_m\mu_m}{2}, \frac{\alpha_m}{2}\big) \big\}$, and $ V_6 = \big\{ - \big\} : \big\{ \big(\frac{\phi}{\alpha}-\mu ,1 \big),  \big(0,1\big), \big(\frac{-\alpha\mu }{2},\frac{\alpha}{2}\big) \big\}:\big\{\big(\frac{\alpha_m\mu_m}{2}, \frac{\alpha_m}{2}\big), \big(1+\frac{\alpha_m\mu_m}{2}, \frac{\alpha_m}{2}\big)\big\}$.

We use \cite[Th. 1.7, 1.11]{Kilbas_2004} and compute  residues of \eqref{cdf_relay_fox} for both contours $L_1$ and $L_2$ at poles $s_1=0$, $\frac{\phi}{\alpha}-\mu $, $-\mu $ and $s_2=0$, $\frac{\alpha s_1+\alpha\mu -\alpha_m\mu_m}{\alpha_m}$ to express the CDF in the high SNR regime, as presented in \eqref{eq:outage_asymptotic}.

We use \eqref{cdf_relay_fox} in \eqref{eq:ber_eqn} with the inner integral $\int_{0}^{\infty}{\gamma}^{\frac{\alpha_m\mu_m-\alpha_ms_2}{2}}\gamma^{p-1} e^{-q\gamma}d{\gamma}$ \cite[3.381/4]{Gradshteyn}, and apply the definition of Fox's H-function \cite[1.1]{Mittal_1972} to get the average BER for the considered hybrid system  as
\begin{flalign}\label{ber_relay_fox}
	\bar{P}_e^{\rm MRC} \approx & \frac{A A_m \phi S_0^{-\alpha\mu }  {\bar{\gamma}_{\rm THz}}^{-{\frac{\alpha\mu }{2}}} C^{\frac{\alpha\mu }{2}}{\bar{\gamma}_{\rm RF}}^{-\frac{\alpha_m\mu_m}{2}}q^{-\big(\frac{\alpha_m\mu_m}{2}+p\big)}}{8 \Gamma(p)\alpha}\nonumber\\&H_{1,0:1,3:2,3}^{0,1:3,0:2,1}\left[\frac{B{C}^{\frac{\alpha}{2}}}{S_0^{\alpha} {\bar{\gamma}_{\rm THz}}^{\frac{\alpha}{2}}}, \frac{ {\bar{\gamma}_{\rm RF}}^{\frac{\alpha_m}{2}} q^{\frac{\alpha_m}{2}}} {B_m}  \Bigg| \begin{matrix} ~~U_7~~ \\ ~~V_7~~  \end{matrix} \right]
\end{flalign}
where $U_7=\big\{\big(1-\frac{\alpha\mu -\alpha_m\mu_m}{2};\frac{\alpha}{2},\frac{\alpha_m}{2} \big)\big\} : \big\{\big(1+\frac{\phi}{\alpha}-\mu ,1 \big)\big\} : \big\{\big(1,1\big),\big(1+\frac{\alpha_m\mu_m}{2}, \frac{\alpha_m}{2}\big) \big\}$, and $ V_7 = \big\{-\big\} : \big\{\big(\frac{\phi}{\alpha}-\mu ,1\big), \big(0,1\big),\big(\frac{-\alpha\mu }{2},\frac{\alpha}{2}\big)\big\} : \big\{\big(\frac{\alpha_m\mu_m}{2},\frac{\alpha_m}{2}\big),\big(p+\frac{\alpha_m\mu_m}{2},\frac{\alpha_m}{2}\big)\big(1+\frac{\alpha_m\mu_m}{2},\frac{\alpha_m}{2}\big)\big\}$.

Finally, we can approximate the ergodic capacity of the hybrid system consisting of RF-MRC and THz  using \eqref{pdf_relay_fox} in \eqref{eq:capacity_eqn}, and apply the similar procedure used for driving \eqref{eq:capacity_sc_af}  to get
\begin{flalign}\label{capacity_relay_fox}
	\bar{\eta}^{\rm MRC} \approx  &  \frac{ A A_m \phi S_0^{-\alpha\mu }  {\bar{\gamma}_{\rm THz}}^{-{\frac{\alpha\mu }{2}}} C^{\frac{\alpha\mu }{2}} {\bar{\gamma}_{\rm RF}}^{-\frac{\alpha_m\mu_m}{2}} } {4\alpha {\rm log}(2)} \nonumber \\ & H_{1,0:1,3:3,3}^{0,1:3,0:2,2} \Bigg[ \frac{B {C}^{\frac{\alpha}{2}}}{S_0^{\alpha} {\bar{\gamma}_{\rm THz}}^{\frac{\alpha}{2}}}, \frac{ {\bar{\gamma}_{\rm RF}}^{\frac{\alpha_m}{2}}} {B_i}  \bigg| \begin{matrix}~~U_8~~	  \\ ~~V_8~~   \end{matrix}    \Bigg]
\end{flalign}
where $U_8 = \big\{\big(1-\frac{\alpha\mu -\alpha_m\mu_m}{2};\frac{\alpha}{2},\frac{\alpha_m}{2} \big)\big\} : \big\{ \big(1+\frac{\phi}{\alpha}-\mu ,1 \big)\big\} : \big\{\big(1,1\big), \big(\frac{\alpha_m\mu_m}{2}, \frac{\alpha_m}{2}\big), \big(1+\frac{\alpha_m\mu_m}{2}, \frac{\alpha_m}{2}\big) \big\}$, and $ V_8= \big\{ - \big\} : \big\{ \big(\frac{\phi}{\alpha}-\mu ,1 \big),  \big(0,1\big), \big(\frac{-\alpha\mu }{2},\frac{\alpha}{2}\big) \big\}:\big\{\big(\frac{\alpha_m\mu_m}{2}, \frac{\alpha_m}{2}\big), \big(\frac{\alpha_m\mu_m}{2}, \frac{\alpha_m}{2}\big), \big(1+\frac{\alpha_m\mu_m}{2}, \frac{\alpha_m}{2}\big)\big\} $.

\begin{table}[t] 
	\caption{List of Simulation Parameters} 
	\label{tab:simulation_parameters} 
	\centering 
	\begin{tabular}{c c} 
		\hline\hline \\  
		\textbf{Parameter}  & \textbf{Value} \\ [1ex] 
		\hline  
		Number of RF antennas ($N$) & $1$-$20$ \\
		$d_{\rm RF}$ & $50$\mbox{m}-$150$\mbox{m}\\
		$d_{\rm THz}$ & $50$\mbox{m}\\
		Absorption coefficient (THz) & $\kappa=2.8 \times 10^{-4}$\\
		THz carrier frequency ($f_{\rm THz}$) & $ 275 $ \mbox{GHz} \\ 
		RF carrier frequency ($f_{\rm RF}$) & $ 800 $ \mbox{MHz} \\ 
		THz	antenna gain ($G_{\rm THz}$)  & $ 55 $ \mbox{dBi} \\
		RF	antenna gain ($G_{\rm RF}$)&  $ 25 $ \mbox{dBi} \\
		$\alpha$-parameter & $1$-$2$ \\ 
		$\mu$-parameter & $0.8$-$2.6$ \\
		$\Omega$-parameter & $1$ \\ 
		$\phi$ & $4.11-231$ \\
		$S_0$ & $0.054$\\
		Jitter standard deviation  ($\sigma_s$) & $2$\mbox-$15$\mbox{cm}\\
		Transmit power & $0$-$40$ \mbox{dBm}\\
		Noise Power (RF) & $ -101$ \mbox{dBm} \\
		Noise Power (THz) & $ -74 $ \mbox{dBm} \cite{Sen_2020_Teranova} \\
		$p,q$ & $0.5,1$ \\
		\hline \hline
	\end{tabular}	
\end{table}

\section{Simulation Analysis}\label{sec:sim_results}
In this section,  we demonstrate the performance of the hybrid RF-THz system and validate the derived analytical results using  Monte-Carlo simulations. We analyze the performance of the considered system when multi-antenna MRC, SC, and EGC diversity techniques are applied. The simulation parameters are provided in Table \ref{tab:simulation_parameters}. We use the 3GPP standard path loss model $L_{\rm RF}({\rm dB}) = 32.4+17.3\log_{10}(d_{\rm RF})+20\log_{10} (10^{-9}f_{\rm RF})$ to compute the path gain for the RF link.  Further, we use the absorption model  $\frac{cG_{\rm THz}}{4\pi f_{\rm THz} d_{\rm THz}} \exp(-\frac{1}{2}\kappa d_{\rm THz})$ (where $c$ denotes the speed of light) for computing the path loss for the THz transmission.  We adopt  \cite{Farid2007} for generating the pointing errors. The $\alpha$-$\mu$ fading channel is generated by appropriately adjusting the MATLAB function "gamrnd". To evaluate the bivariate Fox's H-function, we use MATLAB implementation, as given in \cite{Illi_2017}. 

\begin{figure*}[t]
	\subfigure[]
	{\includegraphics[width=\columnwidth]{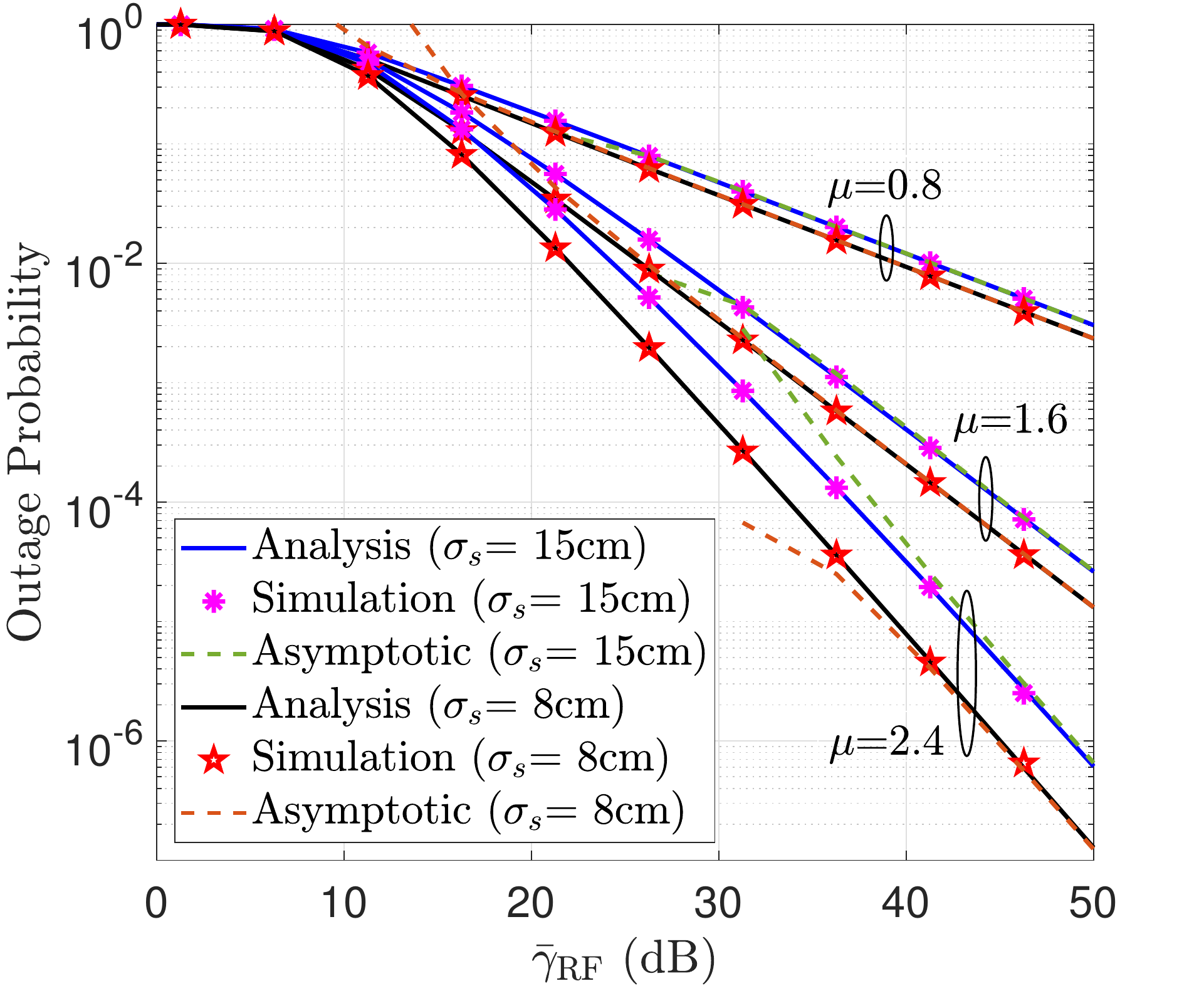}}
	\subfigure[]
	{\includegraphics[width=\columnwidth]{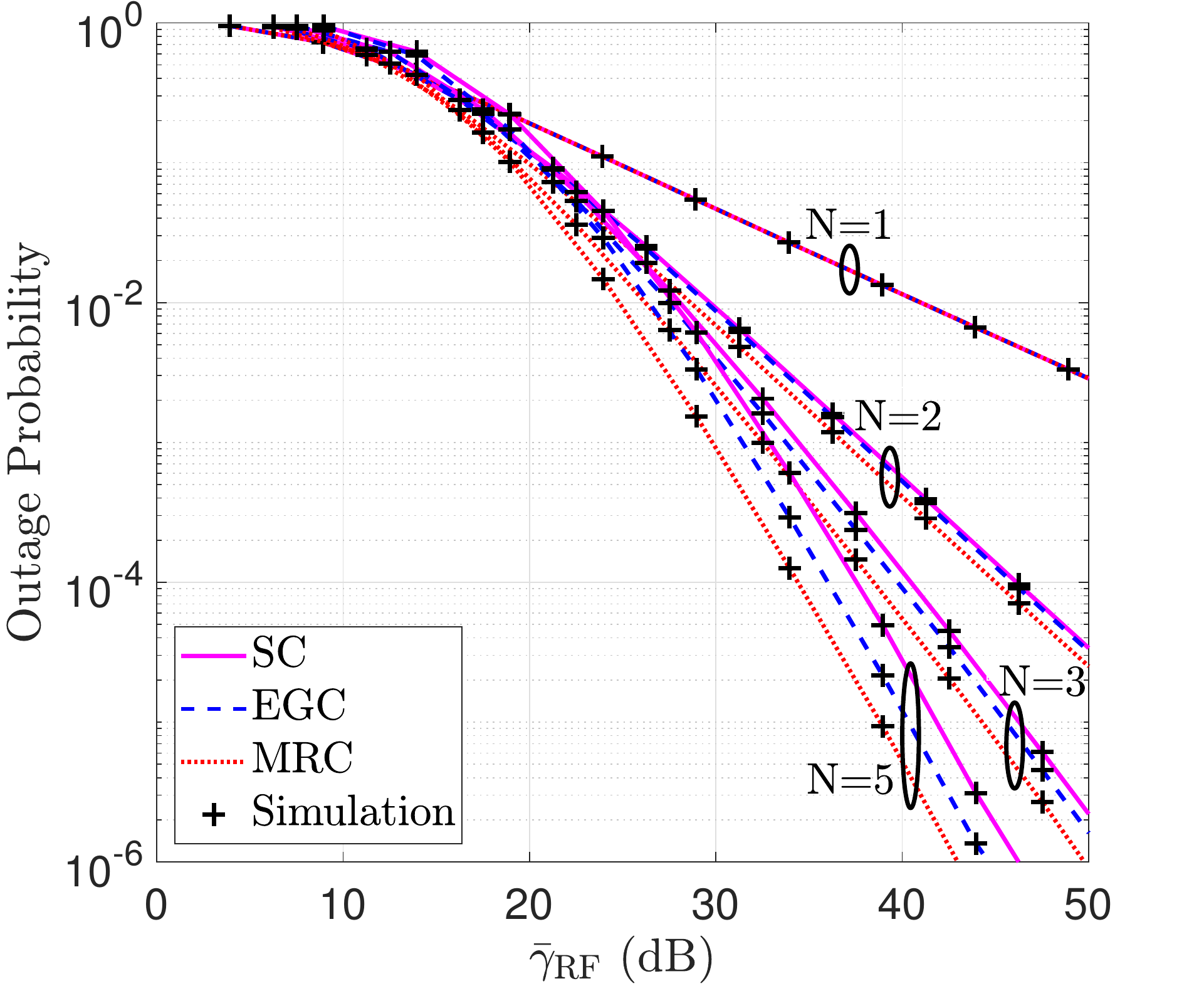}} 
	\caption{Outage probability of hybrid wireless link (a) different $\mu $ and $\sigma_s$ with $N=1$ (b) multi-antenna RF receiver with $\alpha_i=1$ and $\mu_i=1.2$, $\forall i$. }
	\label{fig:outage}
\end{figure*}

\begin{figure*}[t]
	\subfigure[]{\includegraphics[width=\columnwidth]{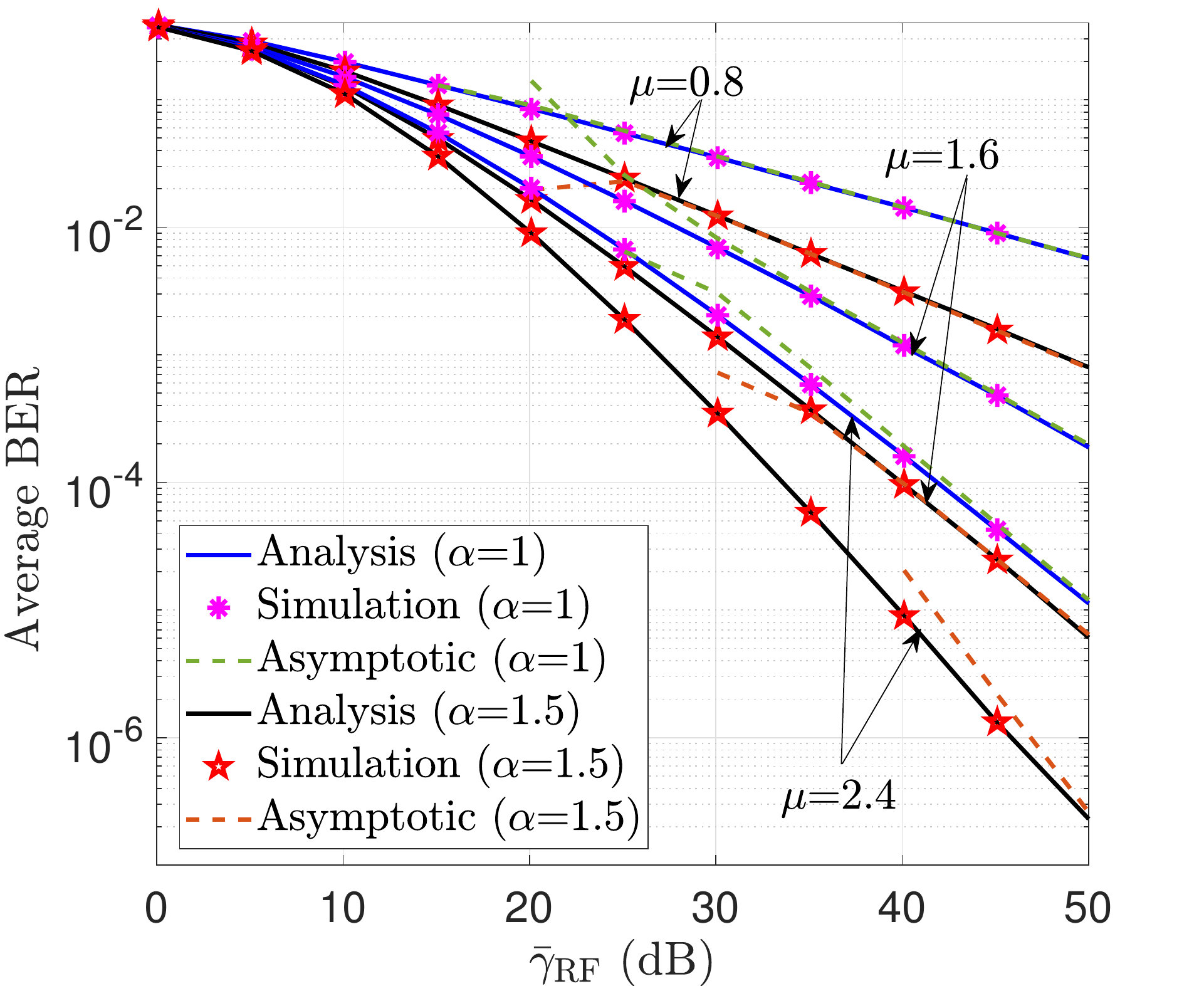}} 
	\subfigure[] {\includegraphics[width=\columnwidth]{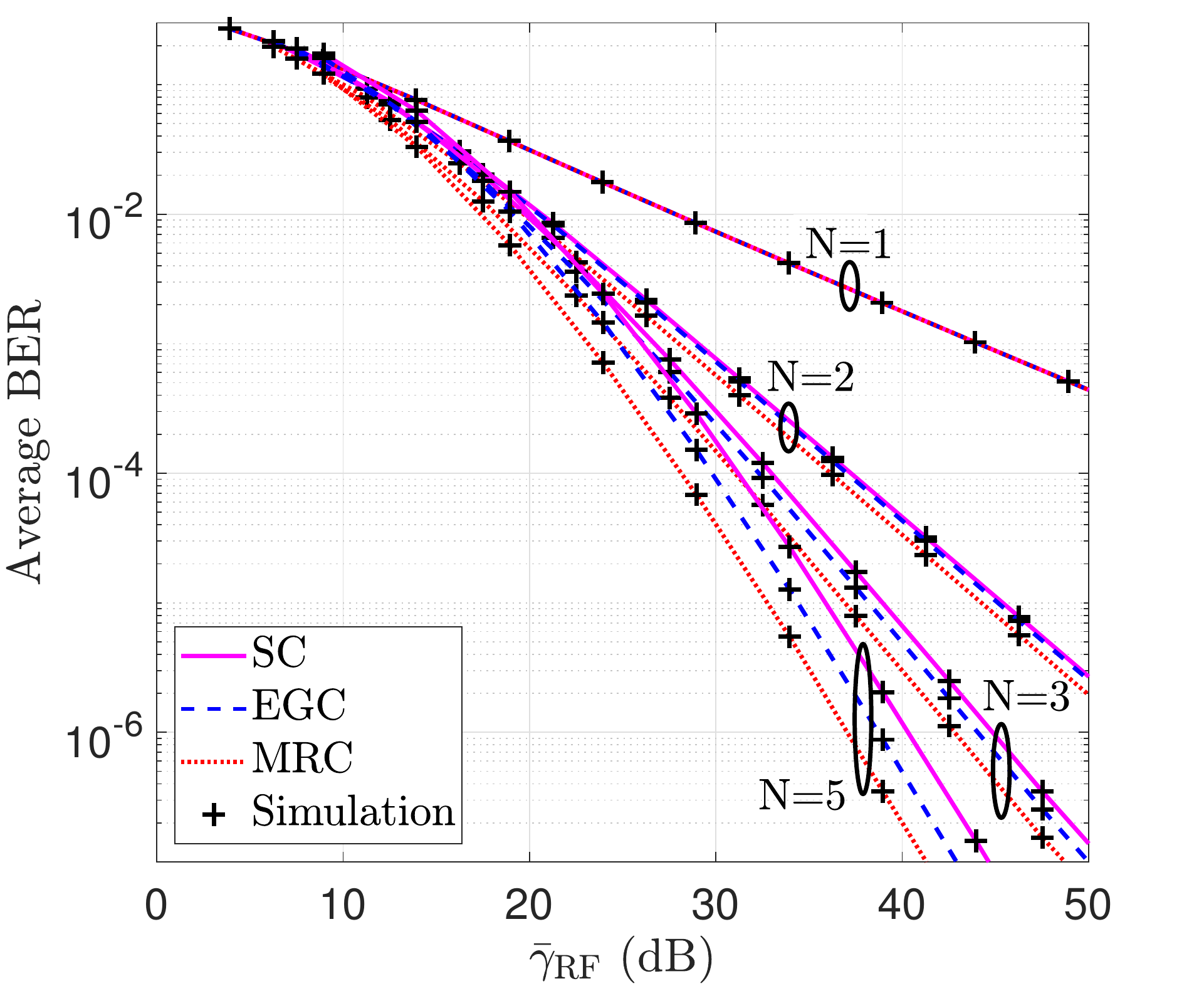}}
	\caption{Average BER performance of hybrid wireless link (a) different $\alpha$ and $\mu$ with $N=1$ (b) multi-antenna RF  with $\alpha_i=1$ and $\mu_i=1.2$ $\forall i$.}
	\label{fig:ber}
\end{figure*}

\begin{figure*}[t]
	\subfigure[] {\includegraphics[width=\columnwidth]{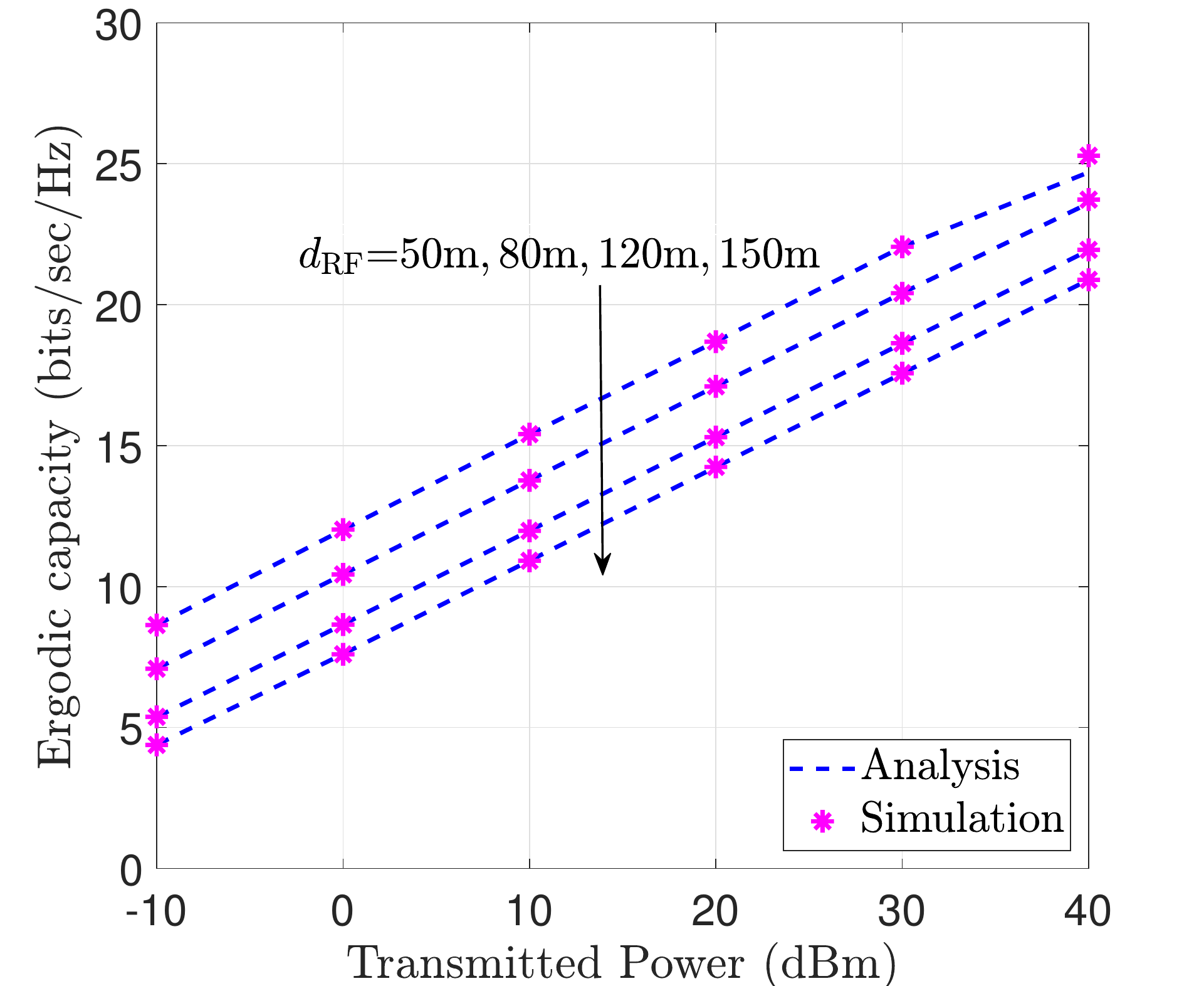}}	
	\subfigure[] {\includegraphics[width=\columnwidth]{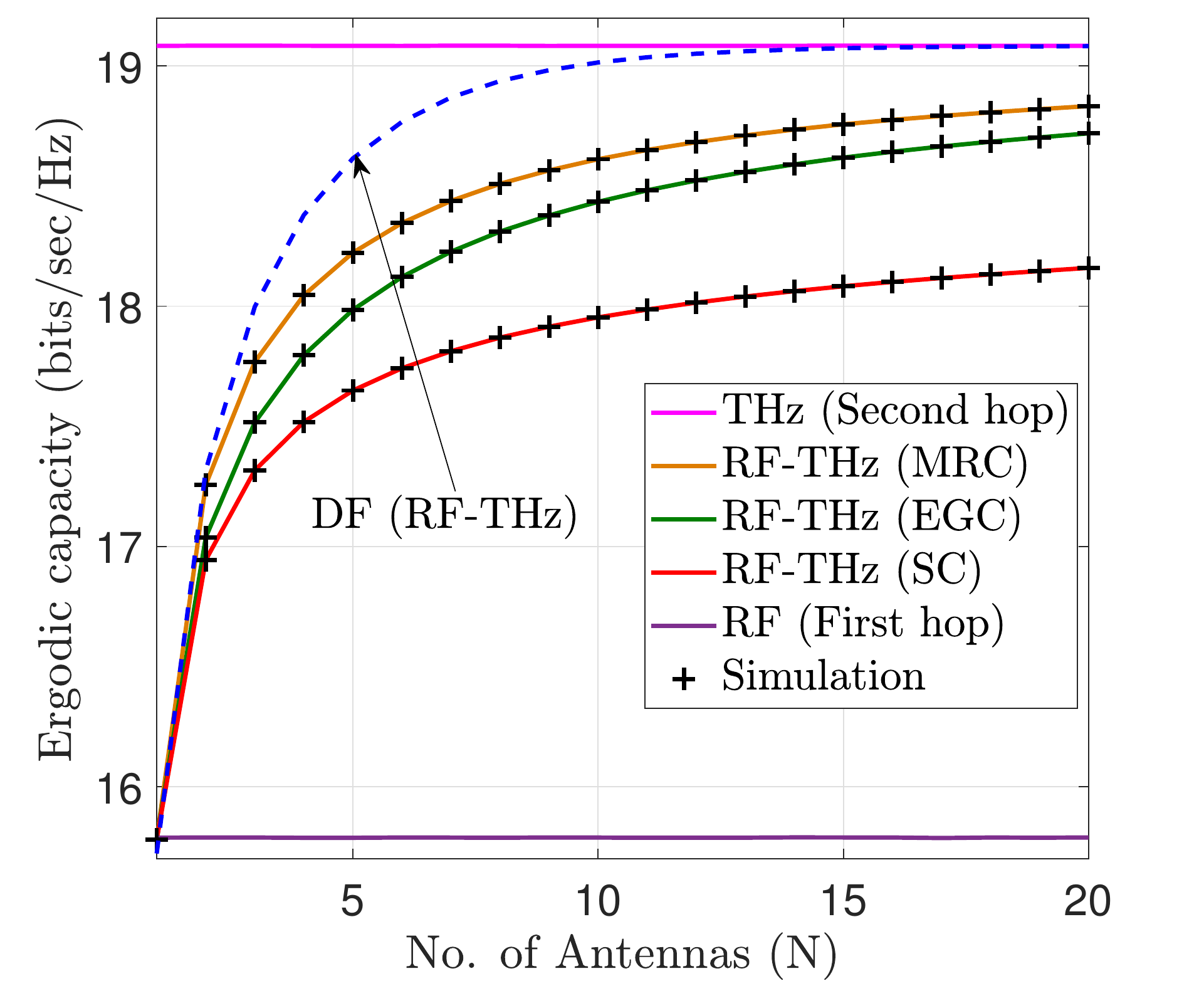}}
	\caption{Ergodic capacity of hybrid wireless link (a) at different RF link distance $d_{\rm RF}$ (b) multi-antenna RF receiver with link distance $d_{\rm RF} = 100$\mbox{m}.}
	\label{fig:capacity}
\end{figure*}
First, we demonstrate the outage probability of the considered hybrid link  over different system and channel configurations, as depicted in Fig. \ref{fig:outage}. In \ref{fig:outage}a, we consider $N=1$ to show the effect of multi-path clustering parameter $\mu $ and pointing errors parameter $\sigma_s$ on the outage probability by considering $d_{\rm RF} = 100 \mbox{m}$, $d_{\rm THz} = 50 \mbox{m}$, $\alpha = 1.5, \alpha_1=2, \mu_1 = 2.5$, and $\gamma_{\rm th} = 4 \mbox{dB}$. The figure illustrates that the outage probability reduces with an increase in  $\mu $ since  clustering improves the channel. Further, the impact of jitter is marginal at a lower $\mu =0.8$ and low RF average SNR but requires almost $4$\mbox{dB} additional SNR if $\sigma_s$ is increased from $8$\mbox{cm} to $15$\mbox{cm} at a higher $\mu =2.4$ and outage probability $10^{-6}$. Since $\phi=14.4788$ for $\sigma_s=8$\mbox{cm} and $\phi=4.1184$ for $\sigma_s=15$\mbox{cm}, and $\alpha_1\mu_1= 5$, the diversity order for different $\mu $ is determined from  the first link as $M= 0.6, 1.2, 1.8$, which can be observed from the slope of corresponding plots.  Moreover, it can be seen that there is no change in the slope for the same $\mu $ even with different $\sigma_s$ since the diversity order is independent of pointing errors for the considered scenario. 

In Fig. \ref{fig:outage}(b), we demonstrate the use of multiple antennas ($N\geq 1$) at the RF to improve the overall performance of the hybrid system. Here,  we compare the performance of a single-antenna system ($N=1$) with a multiple antenna RF system ($N=2$, $N=3$, and $N=5$) using different diversity methods. We take system parameters ($\alpha=2$, $\mu =2.6$, $\sigma_s=10 \mbox{cm}$, $d_{\rm RF}=100$\mbox{m}, and $d_{\rm THz}=50$\mbox{m}) such that the THz link is  stronger (in average sense) than the RF with  $N=1$. It can be observed from Fig. \ref{fig:outage}(b) that the outage performance improves significantly with an increase in the number of antennas achieving a higher gain with $N=2$ system. The outage probability reduces by $30$ times when $N$ is increased from $1$ to $2$  and by $5$ times when $N$ is increased from $2$ to $3$   at an average RF SNR of $40$ \mbox{dB}.  Further, the cumulative gain decreases with an increase in the number of antennas advocating its limited usage.  The performance of the hybrid system depends predominantly on the weaker of both (RF or THz) links, and thus an increase in the number of antennas at the RF link achieves a similar channel quality compared with the  THz saturating the overall performance. Conventionally, the multi-antenna reception for a single link suffers from this limitation due to the channel correlation with an increase in the number of antennas. 

Next, we illustrate the average BER performance of the hybrid system in Fig. \ref{fig:ber}. The effect of non-linearity parameter $\alpha$ of the THz fading channel on the average BER performance  with $N=1$, $d_{\rm RF} = 100 \mbox{m}$, $d_{\rm THz} = 50 \mbox{m}$, $\sigma_s=8 \mbox{cm}$,  $\alpha_1=2.5$, and $\mu_1 = 1.8$ is depicted in. Fig. \ref{fig:ber}(a). The THz link parameters $\alpha$ and $\mu $ effects the average BER performance of the hybrid system significantly. The average BER reduces $100$ fold with an increase in $\alpha$ from $1$ to $1.5$ at an average SNR of $40$ \mbox{dB} and $\mu =2.4$ for an average BER of $10^{-5}$. Similar to the outage probability, the diversity order depends on the first link with $M= 0.4, 0.8, 1.2$ when $\alpha=1$ and $M= 0.6, 1.2, 1.8$ when $\alpha=1.5$. Thus, the diversity for $\mu =2.4$ with $\alpha=1$ and  $\mu =1.6$ with $\alpha=1.5$ are the same, as confirmed in Fig. \ref{fig:ber}(a) (see fourth and fifth plots from the top). In Fig. \ref{fig:ber}(b), we demonstrate the impact of multiple antenna assisted RF reception on the hybrid system with $\alpha=2$, $\mu =2.6$, and $\sigma_s=10 \mbox{cm}$. It can be seen that an average BER of  $5\times 10^{-4}$ can be achieved with a gain of $16$ \mbox{dB} SNR when the number of antennas is increased from $N=1$ to $N=2$. Further, we can find similar observations on  the impact of multiple antenna for the average BER, as inferred from the outage probability plots in Fig. \ref{fig:outage}(b). 

Finally, In Fig. \ref{fig:capacity}, we present the ergodic capacity performance of the two-tier system. In Fig. \ref{fig:capacity}(a),  we demonstrate the ergodic performance with $N=1$ at varying distances of the RF link $d_{\rm RF}=50$\mbox{m} to $d_{\rm RF}=150$\mbox{m} with a fixed THz link distance $d_{\rm THz}=50$\mbox{m} with $\alpha = 2, \mu  = 2.2, \alpha_1=1.5, \mu_1 = 1.8$, and $\sigma_s=10.6 \mbox{cm}$. It can be observed from the figure that the ergodic capacity reduces with an increase in the RF link distance, however, the degradation effect reduces for longer links. In Fig. \ref{fig:capacity}(b), we quantify the number of antennas to achieve the performance equivalent to the THz link, which is stronger (on the average) than the RF with $d_{\rm RF} = 100 \mbox{m}$, $d_{\rm THz}=50$\mbox{m} with $\alpha=2$, $\mu =2.6$, $\alpha_i=1.5$, $\mu_i=1$, $\forall i$, and $\sigma_s=2\mbox{cm}$. When $N=1$, the ergodic capacity of the RF link is almost  $3$\mbox{bits/sec/Hz} lesser than the THz, and thus the hybrid system achieves the RF performance. The figure shows the scaling of the hybrid system with the number of antennas attaining performance close to the THz link. There is a marginal increase in the ergodic capacity of the AF-assisted hybrid system when the number of antennas is increased beyond $N=15$ due to the noise folding of fixed-gain relaying. To demonstrate this, we also plot the DF-based hybrid system with the MRC applied for the RF link, which attains the THz performance within $N=15$ antennas, and saturates thereafter.

It can be seen from Fig.~\ref{fig:outage}, Fig.~\ref{fig:ber}, and Fig.~\ref{fig:capacity} that the derived analytical expressions are in good agreement with Monte Carlo simulations and asymptotic results attain the exact at a relatively lower SNR. Further, the approximation involved in the analysis of MRC is quite accurate since a single $\alpha$-$\mu$ variate can accurately approximate the sum of $\alpha$-$\mu$ variates \cite{Costa_2008_alpha_mu_sum}. Figures also demonstrate the characteristics of different diversity combining techniques: the MRC is optimal, the SC is  inferior to MRC but marginally with a simpler implementation, and  the  performance of the EGC lies between the SC and the MRC. 

\section{Conclusion}\label{sec:conc}
In this paper, we analyzed the performance of a hybrid system consisting of multi-antenna RF in the access link and THz in the backhaul integrated through the fixed-gain AF relaying. We derived the PDF and CDF of the end-to-end SNR of the hybrid system by employing both SC and MRC diversity receivers considering $\alpha$-$\mu$ channel fading for both links along with pointing errors in the THz link. We presented exact, approximation, and asymptotic analysis on the outage probability, average BER, and ergodic capacity of the considered system. We also provided the diversity order of the system, which shows that the performance of the hybrid system can be made independent from RF fading and pointing errors by employing a sufficient number of antennas for RF reception and a higher beam-width for THz transmissions. We provided simulation results that depict a significant gain in the performance with an increase in the number of antennas.  The proposed analysis provided design criteria for the RF-THz system deployment and precludes the use of excessive antennas at the RF since the hybrid system can not exceed the performance of stronger of both links, considered to be the THz link in our simulation environment.  This work can be extended by employing multiple-input and multiple-output (MIMO) system for both RF and THz communications.

\section*{Appendix A}
First, we express the PDF and CDF of the RF link in a compatible form so that statistical results of the combined channel can be obtained in a simplified form. Substituting \eqref{eq:cdf_rf} in \eqref{eq:cdf_max_eqn} we can get the CDF of the first link as
\begin{eqnarray} \label{eq:cdf_max_eqn_sc}
	F_{\gamma_{RF}} (\gamma) = \Bigg[1-\Bigg(\frac{\Gamma\Big(\mu_i, B_i \big(\sqrt{{\gamma}/{\bar{\gamma}_{\rm RF}}}\big)^{\alpha_i}\Big)}{\Gamma (\mu_i)}\Bigg)\Bigg]^N
\end{eqnarray}
Applying  the binomial expansion and using the series expansion of incomplete Gamma function $\Gamma(\mu_i,at) = (\mu_i-1)! \exp(-at) \sum_{j=0}^{\mu_i-1}\frac{(at)^j}{j!}$, we express \eqref{eq:cdf_max_eqn_sc} as
\begin{flalign} \label{eq:cdf_sc_eqn}
	F_{\gamma_{\rm RF}} (\gamma) =& (-1)^k \sum_{k=0}^{N} \Big(\begin{matrix} N \\ k\end{matrix}\Big)  \exp\Big(\hspace{-2mm}-kB_i\sqrt{\frac{\gamma}{\bar{\gamma}_{\rm RF}}}^{\alpha_i}\Big) \nonumber \\ & \times \Bigg(\sum_{j=0}^{\mu_i-1}\frac{B_i\Big(\sqrt{\frac{\gamma}{\bar{\gamma}_{\rm RF}}}\Big)^{j\alpha_i}}{j!}\Bigg)^{k}
\end{flalign}
Applying the multinomial expansion $ (x_1+x_2+\cdots+x_m)^n =  \sum_{k_1+k_2+\cdots+k_m=n}^{} \Big(\begin{matrix}n \\ k_1, k_2,\cdots,k_m \end{matrix}\Big) \prod_{t=1}^{m}x_t^{k_t} $, where $\Big(\begin{matrix}n \\ k_1, k_2,\cdots,k_m \end{matrix}\Big) = \frac{n!}{k_1!k_2!\cdots k_m!}$ on the last term of  \eqref{eq:cdf_sc_eqn}, we get the CDF of the first link  as
\begin{flalign} \label{eq:cdf_snr_rf_sc}
	F_{\gamma_{\rm RF}} (\gamma) =& (-1)^k \sum_{k=0}^{N} \Big(\begin{matrix} N \\ k\end{matrix}\Big) \sum_{k_0+k_1+\cdots+k_{\mu_i-1}=k}^{}  \nonumber\\& \times \Big(\begin{matrix} k \\ k_0+k_1+\cdots+k_{\mu_i-1}\end{matrix}\Big)    \exp\Big(-kB_i\Big({\frac{\gamma}{\bar{\gamma}_{\rm RF}}}\Big)^{\frac{\alpha_i}{2}}\Big)  \nonumber\\&  \times  \prod_{t=o}^{\mu_i-1} \frac{B_i^{k_t} {\gamma}^{(\sum_{t=0}^{\mu_i-1}\frac{t\alpha_ik_t}{2})}}{(t!)^{k_t} {\bar{\gamma}_{\rm RF}}^{(\frac{t\alpha_ik_t}{2})}}    
\end{flalign}
We use  \eqref{eq:pdf_hf_rf} and \eqref{eq:cdf_rf} in  $ f_{\gamma_{\rm RF}} (\gamma)  = N (1 - F_{\gamma_i} (\gamma))^{N-1} f_{\gamma_i} (\gamma) $ to get the PDF of the first link:
\begin{flalign} \label{eq:pdf_max_eqn_sc}
	f_{\gamma_{\rm RF}} (\gamma) =& \Bigg[\Bigg(\frac{\Gamma\Big(\mu_i, B_i\big(\sqrt{\frac{\gamma}{\bar{\gamma}_{\rm RF}}}\big)^{\alpha_i}\Big)}{\Gamma (\mu_i)}\Bigg)\Bigg]^{N-1} \frac{N A_{i} \gamma^{\frac{\alpha_i\mu_i}{2}-1}}{ 2 {\bar{\gamma}_{\rm RF}}^{\frac{\alpha_i\mu_i}{2}} } \nonumber \\ & \times  \exp \Big(- \frac {\mu_i}  {\Omega^{\alpha_i}} {\Big(\sqrt{\frac{\gamma}{\bar{\gamma}_{\rm RF}}}\Big)^{\alpha_i}}\Big)
\end{flalign}
Thus, we use the series expansion of incomplete Gamma function and multinomial expansion in  \eqref{eq:cdf_max_eqn_sc} to get the PDF of the RF link as
\begin{flalign} \label{eq:pdf_snr_rf_sc}
	f_{\gamma_{\rm RF}}(\gamma)=&\frac{N A_i}{2{{\bar{\gamma}_{\rm RF}}^{\frac{\alpha_i\mu_i}{2}}}}\sum_{k_0+k_1+\cdots+k_{\mu_i-1}=N-1}^{}  \nonumber \\ & \times \Big(\begin{matrix} N-1 \\ k_0+k_1+\cdots+k_{\mu_i-1}\end{matrix}\Big) \exp\Big(-\frac{N{{\gamma}}^{\frac{\alpha_i}{2}} B_i}{{{\bar{\gamma}_{\rm RF}}}^{\frac{\alpha_i}{2}}}\Big)   \nonumber \\ &  \times  \prod_{t=o}^{\mu_i-1}\frac{B_i^{k_t}{\gamma}^{(\sum_{t=0}^{\mu_i-1}\frac{t\alpha_ik_t}{2}+\frac{\alpha_i\mu_i}{2}-1)}}{(t!)^{k_t}{\bar{\gamma}_{\rm RF}}^{(\frac{t\alpha_ik_t}{2})}}
\end{flalign}	
Next, we derive the PDF of the mixed link by substituting \eqref{eq:pdf_hfp} and \eqref{eq:pdf_snr_rf_sc} in  \eqref{eq:pdf_eqn_af} using the integral representation of Meijer's-G function \cite{Mathai_2010}:
\begin{flalign} \label{eq:pdf_snr_af_int}
	f_{\gamma}^{\rm SC} (z) = & \frac{NAA_i\phi S_0^{-\alpha\mu } }{ 2 \alpha^2 {{\bar{\gamma}_{\rm RF}}^{\frac{\alpha_i\mu_i}{2}}} {\bar{\gamma}_{\rm THz}}^{\frac{\alpha\mu}{2}}} \sum_{k_0+k_1+\cdots+k_{\mu_i-1}=N-1}^{}  \nonumber \\ & \times \Big(\begin{matrix} N-1 \\ k_0+k_1+\cdots+k_{\mu_i-1}\end{matrix}\Big)  \nonumber \\ & \times \prod_{t=o}^{\mu_i-1} \frac{B^{k_t} (z)^{(\sum_{t=0}^{\mu_i-1}\frac{t\alpha_i k_t}{2}+\frac{\alpha_i\mu_i}{2}-1)}}{(t!)^{k_t} {\bar{\gamma}_{\rm RF}}^{(\frac{t\alpha_ik_t}{2})}}  \nonumber \\ & \times \frac{1}{2\pi \J} \int_{\mathcal{L}_1} \Gamma(0-S_1) \bigg( \frac{Nz^{\frac{\alpha_i}{2}}B_i}{{{\bar{\gamma}_{\rm RF}}}^{\frac{\alpha_i}{2}}}\bigg)^{S_1} dS_1 \nonumber \\ & \times \frac{1}{2\pi \J} \int_{\mathcal{L}_2} \frac{\Gamma(-\mu+\frac{\phi}{\alpha}-S_2) \Gamma(-S_2) \Gamma(\mu+S_2) }{\Gamma(1-\mu+\frac{\phi}{\alpha}-S_2) \Gamma(1+\mu+S_2)}  \nonumber \\ & \times \bigg(\frac{B}{S_o^{\alpha} {\bar{\gamma}_{\rm THz}}^{\frac{\alpha}{2}}}\bigg)^{S_2} dS_2 I_{1}
\end{flalign}
where $\mathcal{L}_1$ and $\mathcal{L}_2$ are the contours of line integrals. The inner integral $I_{1}$ is solved by applying \cite[3.194/3]{Gradshteyn} and \cite[8.384/1]{Gradshteyn}:
\begin{flalign}
	&I_{1} = \int_{0}^{\infty}  x^{(\frac{\alpha\mu}{2}+\frac{\alpha S_2}{2}-1)}   \big(\frac{x + C}{x}\big)^{\sum_{t=0}^{\mu_i-1}\frac{t\alpha_i k_t}{2}+\frac{\alpha_i\mu_i}{2}+\frac{\alpha_iS_1}{2}}dx = \nonumber \\ & \frac{C^{(\frac{\alpha\mu+\alpha S_2}{2})}  \Gamma\big(\frac{-\alpha\mu-\alpha S_2}{2}\big)\Gamma\big(\frac{\alpha\mu-\sum_{t=0}^{\mu_i-1}{t\alpha_ik_t-\alpha_i\mu_i-\alpha_iS_1+\alpha S_2}}{2}\big)}{\Gamma\big(-\sum_{t=0}^{\mu_i-1} \frac{t\alpha_ik_t}{2}-\frac{\alpha_i\mu_i}{2}-\frac{\alpha_iS_1}{2}\big)}
\end{flalign}
Substituting $I_{1}$ back into \eqref{eq:pdf_snr_af_int}, converting $S_1\to-S_1$ and applying the definition of Fox's H-function \cite[1.1]{Mittal_1972}, we get the PDF of multiantenna RF mixed with THz employing MRC in \eqref{eq:pdf_snr_af}. The CDF can be derived using $F_{\gamma}^{\rm SC} (z)=\int_{0}^{\gamma} f_{\gamma}^{\rm SC} (z) dz$
\begin{flalign} 
	&F_{\gamma}^{\rm SC} (z) =  \frac{NAA_i\phi S_0^{-\alpha\mu } C^{\frac{\alpha\mu}{2}} }{ 2 \alpha^2 {{\bar{\gamma}_{\rm RF}}^{\frac{\alpha_i\mu_i}{2}}} {\bar{\gamma}_{\rm THz}}^{\frac{\alpha\mu}{2}}} \sum_{k_0+k_1+\cdots+k_{\mu_i-1}=N-1}^{} \nonumber \\ & \Big(\begin{matrix} N-1 \\ k_0+k_1+\cdots+k_{\mu_i-1}\end{matrix}\Big)   \prod_{t=o}^{\mu_i-1} \frac{B^{k_t} }{(t!)^{k_t} {\bar{\gamma}_{\rm RF}}^{(\frac{t\alpha_ik_t}{2})}}  \nonumber \\ & \frac{1}{2\pi \J} \int_{\mathcal{L}_1} \Gamma(0+S_1) \bigg(\frac{ {{\bar{\gamma}_{\rm RF}}}^{\frac{\alpha_i}{2}}}{NB_i}\bigg)^{S_1} dS_1  \nonumber \\ &  \frac{1}{2\pi \J} \int_{\mathcal{L}_2} \frac{\Gamma(-\mu+\frac{\phi}{\alpha}-S_2) \Gamma(-S_2) \Gamma(\mu+S_2) }{\Gamma(1-\mu+\frac{\phi}{\alpha}-S_2) \Gamma(1+\mu+S_2)} \bigg( \frac{B C^{\frac{\alpha}{2}} }{S_o^{\alpha} {\bar{\gamma}_{\rm THz}}^{\frac{\alpha}{2}}}\bigg)^{S_2} dS_2    \nonumber \\ &     \frac{ \Gamma\big(\frac{-\alpha\mu-\alpha S_2}{2}\big)  \Gamma\big( \frac{\alpha\mu}{2} - \sum_{t=0}^{\mu_i-1}\frac{t\alpha_i k_t}{2}-\frac{\alpha_i\mu_i}{2}+\frac{\alpha_iS_1}{2} +\frac{\alpha S_2}{2}\big)     }{\Gamma\big(-\sum_{t=0}^{\mu_i-1}\frac{t\alpha_i k_t}{2}-\frac{\alpha_i\mu_i}{2}+\frac{\alpha_iS_1}{2}\big)}  \nonumber \\ &  \int_{0}^{\gamma} (z)^{(\sum_{t=0}^{\mu_i-1}\frac{t\alpha_i k_t}{2}+\frac{\alpha_i\mu_i}{2}-\frac{\alpha_iS_1}{2}-1)} dz
\end{flalign}
We can simplify the inner integral $\int_{0}^{\gamma} (z)^{(\sum_{t=0}^{\mu_i-1}\frac{t\alpha_i k_t}{2}+\frac{\alpha_i\mu_i}{2}-\frac{\alpha_iS_1}{2}-1)} dz$ in terms of Gamma function and applying the definition of Fox's H-function \cite[1.1]{Mittal_1972} we obtain the CDF of multiantenna RF mixed with THz employing MRC in \eqref{eq:cdf_snr_af}, which proves Theorem 1.

\section*{Appendix B}
First, we derive the PDF of $\gamma_{\rm RF} = \sum_{i=1}^{N} \gamma_{r_i}$ using the moment generating function (MGF) based approach. We  apply inverse Laplace transform to get the PDF  $f_{\gamma_{\rm RF}}(\gamma) = \mathcal{L}^{-1}\mathcal{M}_\gamma(s)$, where $\mathcal{M}(s)$ is defined as $ \mathcal{M}_{\gamma}(s) = \prod_{i=1}^{N} \mathcal{M}_{\gamma_i}(s) $ with $\mathcal{M}_{\gamma_i}(s) = \int_{0}^{\infty} e^{-s\gamma_i} f_{\gamma_i}(\gamma) d\gamma$. Representing \eqref{eq:pdf_hf_rf} in its Meijer's G equivalent and simplifying, we get 

\begin{eqnarray}\label{eq:mgf_eqn_mrc}
&\mathcal{M}_{\gamma}(s) = \prod_{i=1}^{N} \Bigg[\frac{A_{i}{\bar{\gamma}_{\rm RF}}^{\frac{-\alpha_i\mu_i}{2}}}{2 s^{\frac{\alpha_i\mu_i}{2}}}  \frac{1}{2\pi i} \int_{\mathcal{L}}^{} \Gamma(0-S_1)    \nonumber \\ &  \times \Gamma\Big(\frac{\alpha_i\mu_i+\alpha_iS_1}{2}\Big) \Big(\frac{B_i} {s^{\frac{\alpha_i}{2}}{\bar{\gamma}_{\rm RF}}^{\frac{\alpha_i}{2}}}\Big)^{S_1} dS_1\Bigg]    
\end{eqnarray}

Applying the inverse Laplace transform in \eqref{eq:mgf_eqn_mrc} with standard Mathematical procedure, and using the definition of multivariate Fox's H-function,  we get the PDF of MRC-RF:	
\begin{eqnarray} \label{eq:pdf_rf_MRC}
	f_{\gamma_{\rm RF}}(\gamma)& = \prod_{i=1}^{N} \frac{A_{i}}{2 {\bar{\gamma}_{\rm RF}}^{\frac{\alpha_i\mu_i}{2}}} \bigg(\frac{1}{\gamma}\bigg)^{\big(\sum_{i=1}^{N}\frac{-\alpha_i\mu_i}{2}+1\big)}\nonumber \\ & H_{0,1;1,1;\cdots;1,1}^{0,0;1,1;\cdots;1,1} \left[ \biggl\{\frac{B_i {\gamma}^{\frac{\alpha_i}{2}}} { {\bar{\gamma}_{\rm RF}}^{\frac{\alpha_i}{2}}}\biggr\}_{i=1}^{N} \Bigg| \begin{matrix}~~U_9~~ \\  ~~V_9~~ \end{matrix}  \right]
\end{eqnarray}
where  $U_9 = \bigl\{-\bigr\}: \Bigl\{\Big(1-\frac{\alpha_1\mu_1}{2}, \frac{\alpha_1}{2}\Big),\cdots, \Big(1-\frac{\alpha_i\mu_i}{2}, \frac{\alpha_i}{2}\Big)\Bigr\} $ and $V_9 = \Bigl\{\Big(1-\sum_{i=1}^{N}\frac{\alpha_i\mu_i}{2}; \frac{\alpha_1}{2},\cdots,\frac{\alpha_{r_N}}{2} \Big)\Bigr\}: \Bigl\{(0,1),\cdots,(0,1)_N\Bigr\} $.

Next, we use \eqref{eq:pdf_rf_MRC} and \eqref{eq:pdf_hfp} in \eqref{eq:pdf_eqn_af} with the inner integral $\int_{0}^{\infty}\Big(\frac{(x+C)}{x}\Big)^{\big(\sum_{i=1}^{N}\frac{\alpha_i\mu_i+\alpha_{r_iS_i}}{2}\big)} x^{\frac{\alpha\mu +\alpha \zeta}{2}-1}  dx =  \frac{C^{\frac{\alpha\mu +\alpha \zeta}{2}}\Gamma\big(\frac{-\alpha\mu -\alpha \zeta}{2}\big)\Gamma\big(\sum_{i=1}^{N}\frac{-\alpha_i\mu_i-\alpha_{r_iS_i}}{2} + \frac{\alpha\mu +\alpha \zeta}{2}\big)}{\Gamma\big(\sum_{i=1}^{N}\frac{-\alpha_i\mu_i-\alpha_{r_iS_i}}{2}\big)}$, and changing the sign of integral contour $S_1$, we get
\begin{eqnarray}
&f_{\gamma}^{\rm MRC}(\gamma) = \prod_{i=1}^{N} \frac{A_{r_i} \gamma^{\big(\sum_{i=1}^{N}\frac{\alpha_i\mu_i}{2}-1\big)}  A\phi S_0^{-\alpha\mu } C^{\frac{\alpha\mu}{2}} }{2 {\bar{\gamma}_{\rm RF}}^{\frac{\alpha_i\mu_i}{2}}  \alpha^2 {\bar{\gamma}_{\rm THz}}^{\frac{\alpha\mu }{2}}}    \nonumber \\ &   \Bigg[ \Big(\frac{1}{2\pi \J}\Big)^N  \prod_{i=1}^{N} \int_{\mathcal{L}}^{} \Gamma(0+S_i) \Gamma\Big(\frac{\alpha_i\mu_i-\alpha_iS_i}{2}\Big)   \nonumber \\ & \bigg( \prod_{i=1}^{N} \frac{{\bar{\gamma}_{\rm RF}}^{\frac{\alpha_i}{2}}} { B_i \gamma^{\frac{\alpha_i}{2}}}\bigg)^{S_i} \frac{2\pi i}{\Gamma\big(\sum_{i=1}^{N}\frac{\alpha_i\mu_i-\alpha_iS_1}{2}\big)}  dS_i\Bigg] \nonumber \\ & \frac{1}{2\pi \J}\int_{L_2} \frac{\Gamma(-\mu+\frac{\phi}{\alpha}-\zeta) \Gamma(0-\zeta) \Gamma(\mu+\zeta) }{\Gamma(1-\mu+\frac{\phi}{\alpha}-\zeta) \Gamma(1+\mu+\zeta)} \bigg( \frac{B C^{\frac{\alpha}{2}}}{S_o^{\alpha} {\bar{\gamma}_{\rm THz}}^{\frac{\alpha}{2}}}\bigg)^{\zeta}d\zeta\nonumber\\&\frac{\Gamma\big(\frac{-\alpha\mu -\alpha \zeta}{2}\big)\Gamma\big(\sum_{i=1}^{N}\frac{-\alpha_i\mu_i+\alpha_{r_iS_i}}{2}+\frac{\alpha\mu +\alpha \zeta}{2}\big)}{\Gamma\big(\sum_{i=1}^{N}\frac{-\alpha_i\mu_i+\alpha_{r_iS_i}}{2}\big)}
\end{eqnarray}
We apply the definition of multivariate Fox's H-function \cite[A.1]{Mathai_2010} to get the PDF of the RF link with the MRC mixed with THz in \eqref{eq:pdf_af_MRC}.

Finally, we derive the  CDF $F_{\gamma_{\rm RF}}^{\rm MRC}(\gamma) = \int_{0}^{\gamma} f_{\gamma_{\rm RF}}^{\rm MRC}(\gamma) d\gamma$ using the inner integral $\int_{0}^{\gamma}\gamma^{\big(\sum_{i=1}^{N}\frac{-\alpha_{r_iS_i}+\alpha_i\mu_i}{2}-1\big)}d\gamma=\frac{\gamma^{\big(\sum_{i=1}^{N}\frac{-\alpha_{r_iS_i}+\alpha_i\mu_i}{2}\big)}}{\Gamma\big(\sum_{i=1}^{N}\frac{-\alpha_{r_iS_i}+\alpha_i\mu_i}{2}\big)}$, and applying \cite[A.1]{Mathai_2010} in terms of $N+1$-variate Fox's H-function in \eqref{eq:cdf_af_MRC}, which concludes the proof of Theorem 2.

\bibliographystyle{IEEEtran}
\bibliography{BibTex}

\end{document}